\documentclass[nofootinbib,pre,tighten]{revtex4}

\usepackage{graphicx}
\usepackage{color}

\begin{document}

\title[Polydisperse systems of counterions]{Planar screening by charge polydisperse counterions}

\author{M. Trulsson$^{1,2}$, E. Trizac$^2$, L. \v{S}amaj$^{3}$}

\address{$^1$Theoretical Chemistry, Lund University, Sweden}

\address{$^2$LPTMS, CNRS, Univ. Paris-Sud, Universit\'e Paris-Saclay,
91405 Orsay, France}

\address{$^3$Institute of Physics, Slovak Academy of Sciences,
D\'ubravsk\'a cesta 9, 845 11 Bratislava, Slovakia}


\begin{abstract}
We study how a neutralising cloud of counterions screens the electric
field of a uniformly charged planar membrane (plate), when the counterions
are characterised by a distribution of charges (or valence), $n(q)$. We work
out analytically the one-plate and two-plate cases, at the level of non-linear
Poisson-Boltzmann theory. The (essentially asymptotic) predictions are successfully 
{compared}
to numerical {solutions} of the full Poisson-Boltzmann theory, but also 
{to Monte Carlo simulations}. The counterions with smallest valence control the long-distance features
of interactions, and may qualitatively change the results pertaining to the classic monodisperse
case where all counterions have the same charge. Emphasis is put 
on continuous distributions $n(q)$, for which new power-laws can be evidenced,
be it for the ionic density or the pressure, in the one- and two-plates situations 
respectively. We show that for discrete distributions, more relevant for experiments, these scaling laws {persist in an  
intermediate but {yet} observable range}. Furthermore, it appears that from a practical
point of view, hallmarks of the continuous $n(q)$ behaviour is already featured by discrete mixtures
with a relatively small number of constituents.
\end{abstract}




\maketitle

\section{Introduction}

Polydispersity refers to non-uniformity in some property. For soft matter,
it can pertain to size, surface features, charges, electrolytic content etc.,
and lead to baffling complexity in structural or dynamical properties \cite{ILMR,TaAr95}.
Not only can some structures be destabilised 
\cite{vdLBD13}, nucleation \cite{AuFr01} or compressibility \cite{BCCD11} be suppressed,
but also fractionation may ensue \cite{Bart00,SoWi11} or super-lattices appear \cite{ElMF93,BCGJ16}.
In the limit of a continuous mixture of hard sphere, it was shown 
that optimal packing yields the formation of macroscopic aggregates,
in a scenario that bears similarities with Bose-Einstein condensation \cite{ZBTC99,BlCu01}.

The present paper is devoted to charged fluids, and to the physics of screening by a polydisperse ensemble of counterions,
having different valence. There is a number of reasons for investigating such problems. First of all,
multivalent ions of distinct charges are routinely found in a wealth of situations. One may think here 
of spermine and spermidine ions in biological systems \cite{AnRe80}. Also, the upsurge of interest for nanocolloidal systems 
provides a motivation for our work, where the presence of distinct species with specific charges should be accounted for
\cite{exp1,exp2}.
A specific feature of a linear description, \`a la Debye and H\"uckel \cite{Levin02},
of the kind of polydispersity we are interested in, is entirely subsumed into the so-called Debye length, which is a very
coarse measure of the dispersion in ionic valences. Yet, non-linear effects, overlooked at the Debye-H\"uckel
level, deeply affect the structure of the electric double-layer in the vicinity 
of charged macromolecules \cite{UhGK01,TeTr04,TrTe06}: the sole Debye length is not sufficient
to characterise screening, and thus interactions between charged bodies. 
Our analysis is worked out at the level of the non-linear 
Poisson-Boltzmann theory, where it is interesting to note that the problem of mixed valences
has been investigated in the pioneering paper of Gouy \cite{Gouy10}\footnote{
Gouy remarked that for a uniformly charged plate in an otherwise unbounded electrolyte,
not only 1:1 salt situations, but also 2:1 and 1:2 were solvable analytically 
at Poisson-Boltzmann level \cite{Gouy10}.  Curiously enough, so is the case in cylindrical geometry \cite{TrTe06},
where the key to resolution lies in a mapping to Painlev\'e III equations 
\cite{TrWi97}. For other electrolyte asymmetries, no closed-form solutions can be found.}.
Yet, exact results are scarce, even in the planar geometry to which we restrict
our study. To complement the analytical derivation, Monte Carlo simulation results will also
be reported.

To magnify non-linear effects, we will be interested in a counterion only system (the limit 
of a completely deionised solution \cite{deionised}),
and analyse screening of planar charged bodies. A number of new analytical results can then be derived. 
In this very geometry, parallel like-charged plates interact at long-distances in a universal 
fashion, provided only one type of counterion is present in the solution (mono-disperse case). It can indeed be
readily shown that the corresponding pressure behaves like $P \sim \pi kT/(2 \ell_{\rm B} d^2 )$ 
where $d$ is the inter-plate distance, and $\ell_{\rm B}$ is the Bjerrum length defined below, scaling like the elementary charge squared 
 \cite{Andelman06}.
Thus, the previous large-$d$ result is universal, independent of the charge on the plates.
This result can be generalised to any polydisperse counterionic mixture,
under the proviso that there is a lower bound $q_{\text{min}}>0$ in the valence distribution.
{Here}, the ions with smaller valence are less attracted to the charged plates,
and are those mediating the interaction force. Those ions with valence larger than $q_{\text{min}}$
screen the plates' bare charge, reducing its effective value, which however does not enter the 
large-$d$ behaviour.
We thus expect the universal asymptotic $P \propto  q_{\text{min}}^{-2} d^{-2} $
to be valid as well for a mixture, be it discrete or continuous, as long as
$q_{\text{min}}>0$. We will see in particular that whenever $q_{\text{min}}=0$,
the situation changes completely, and that new power-law regimes emerge,
with a $d$-exponent smaller than 2 that can be tuned continuously. This is a consequence of less efficient
screening, resulting in a severe enhancement of effective interactions.  
In a sense to be specified though, these interactions keep some level of universality. 

The paper is organized as follows. We present in Sections \ref{sec:onewall}
and \ref{sec:twowalls} the results for the one-plate and two-plates
geometries. These are tested against numerical simulations, of two 
distinct types: numerical resolution of Poisson-Boltzmann theory on the one 
hand, and Monte Carlo simulations on the other hand. 
The numerical techniques used are sketched in the appendix.
Finally, our main results are recovered and extended in a heuristic and rather direct way
in section \ref{sec:discussion}. A significant part of the analytical treatment 
(with the notable exception of the statements that do not pertain to asymptotic 
results) is devoted not only to continuous distributions $n(q)$, but furthermore,
to distributions having a vanishing minimum charge $q_{\text min}$. The reason is that 
the behaviour of $n(q)$ for $q\to 0$ is at the root of new scaling laws for the long-distance 
ionic profiles, or interplate pressures. Indeed, those counterions with a large valency 
will be more attracted to the charged plates, while the others are less localised, and 
play a more important role in large scale features.
Yet, the corresponding ``continuous models'' might be viewed as somewhat artificial,
since any physical system exhibiting polydispersity in counterion charge 
will have $q_{\text min}>0$. In section \ref{sec:discussion}, we shall address that legitimate
concern, and show that the newly found power-laws can be observed {over an intermediate
range if} 
$q_{\text min}>0$ or
in discrete systems. In addition, we will present numerical data illustrating the fact 
that in some cases, a small number of species is sufficient for a system to exhibit the
continuous polydispersity asymptotics. Some attention will also be paid to 
universal features that may characterize density profiles and equations of state.

\section{One-wall geometry}
\label{sec:onewall}
We consider a hard wall of dielectric constant $\epsilon'$ localised in 
the half-space $x<0$. The Cartesian $(y,z)$ coordinates are unbounded\footnote{
When performing a mean-field type of analysis, space dimension does not 
play a particular role and up to irrelevant constants, the same
Poisson equation is solved irrespective of dimensionality.}.
The surface of the wall at $x=0$ carries a constant surface charge 
density $\sigma e$ ($e$ is a unit charge and say $\sigma>0$).
Mobile particles, confined in the half-space $x>0$,
are immersed in a medium of dielectric constant $\epsilon$. 
We assume for simplicity that $\epsilon'=\epsilon$, i.e. there are no 
electrostatic image charges.
Particles can have various charges, with sign opposite to that of the
plate: they are counterions. 
Let $\rho(x)$ be the particle charge density (per unit surface of the wall)
at distance $x$ from the wall. 
The condition of overall electroneutrality reads
\begin{equation} \label{neutrality}
\sigma e + \int_0^{\infty} {\rm d}x\, \rho(x) = 0 .
\end{equation} 

The mean electrostatic potential $\psi(x)$ fulfils the Poisson equation
\begin{equation} \label{Poisson}
\frac{{\rm d}^2\psi(x)}{{\rm d}x^2} = - \frac{4\pi}{\epsilon} \rho(x) .
\end{equation}
Integrating this equation over $x$ from $0$ to $\infty$, 
the requirement of electroneutrality (\ref{neutrality}) is consistent 
with the couple of boundary conditions (BCs)
\begin{equation} \label{BCorig}
\psi'(0) = - \frac{4\pi\sigma e}{\epsilon} , \qquad \psi'(\infty) = 0 . 
\end{equation} 

\subsection{Monodisperse case}
We first recapitulate briefly the monodisperse results \cite{Andelman06} where
all mobile ions possess the same charge, say $-e<0$ 
(i.e. their valence is $q=1$). 
Denoting by $n(x)$ the particle number density at $x$, the charge density
is simply $\rho(x) = - e n(x)$.

The statistical mechanics of the system is described
by the mean-field Poisson-Boltzmann (PB) theory \cite{Gouy10,Chapman13},
provided Coulombic coupling is small enough \cite{NJMN05,SaTr11,Rque03}.
In the PB approach, the density of particles at a given point is proportional 
to the corresponding Boltzmann weight of the mean electrostatic potential, 
\begin{equation}
n(x) = f_0 \, {\rm e}^{\beta e \psi(x)} ,
\end{equation}
where $f_0$ is a normalisation constant and $\beta$ denotes the inverse temperature.
Introducing the reduced potential
\begin{equation} \label{reducedpot}
\phi(x) = \beta e \psi(x) , \qquad n(x) = f_0 \, {\rm e}^{\phi(x)} , 
\end{equation}
this mean-field assumption applied to (\ref{Poisson}) leads to the PB equation
\begin{equation} \label{PB}
\frac{{\rm d}^2\phi(x)}{{\rm d}x^2} = 4\pi \ell_{\rm B} f_0 \, {\rm e}^{\phi(x)} ,
\end{equation}
where $\ell_{\rm B}\equiv \beta e^2/\epsilon$ is the Bjerrum length.
Note that the shift of $\phi$ by a constant only renormalizes $f_0$.
We fix the potential gauge by setting 
\begin{equation} \label{gauge}
\phi(0) = 0 
\end{equation}
at the wall. Once a gauge has been chosen, $f_0$ is directly related to the contact density 
of counterions, $n(0)$.
The BCs (\ref{BCorig}) read for the reduced potential as follows
\begin{equation} \label{BCreduced}
\phi'(0) = - 4\pi \ell_{\rm B} \sigma , \qquad \phi'(\infty) = 0 . 
\end{equation} 
Since $\phi'(x)\le 0$ and with regard to the gauge (\ref{gauge}),
it holds that $\phi(x)\le 0$. 
Due to the absence of the neutralising bulk background 
(like in jellium models), the bulk particle density vanishes 
and so $\phi(x)$ goes to $-\infty$ at asymptotically large $x$. 
This is the reason why an approach \`a la Debye-H\"uckel necessarily fails
here, since it relies on linearising the problem around a point 
of reference, taken usually for a one macroion problem as the bulk surrounding 
electrolyte. Here, we have no electrolyte, only counterions.
 
Multiplying the PB equation (\ref{PB}) by $\phi'(x)$, it can be rewritten as
\cite{Andelman06}
\begin{equation} \label{intPB}
\frac{1}{2} \frac{{\rm d}}{{\rm d}x} \left[ \phi'(x) \right]^2
= 4 \pi \ell_{\rm B} f_0 \, \frac{{\rm d}}{{\rm d}x} {\rm e}^{\phi(x)} , \qquad
\frac{1}{2} \left[ \phi'(x) \right]^2 = 4 \pi \ell_{\rm B} f_0 \, {\rm e}^{\phi(x)} , 
\end{equation}  
where the integration constant equals to 0 due to the BCs
$\phi'(x)\to 0$ and ${\rm e}^{\phi(x)}\to 0$ in the limit $x\to\infty$.
The gauge (\ref{gauge}) and the first BC in (\ref{BCreduced}), when considered
in (\ref{intPB}), fix the normalization constant to
$f_0 = 2\pi \ell_{\rm B} \sigma^2$.
The resulting first-order differential equation
\begin{equation}
\phi'(x) = - 4\pi \ell_{\rm B} \sigma {\rm e}^{\phi(x)/2} 
\end{equation}
with the BC $\phi(0)=0$ is solvable by the method of the separation of 
variables:
\begin{equation}
\phi(x) = - 2 \ln \left( 1 + \widetilde{x} \right) ,
\end{equation}
where $\widetilde{x}$ is the dimensionless distance given by
\begin{equation} \label{xtilde}
\widetilde{x}\equiv \frac{x}{\mu} , \qquad \mu=\frac{1}{2\pi\ell_{\rm B} \sigma}
\end{equation}
$\mu$ being the Gouy-Chapman length.
The electric potential goes to $-\infty$ at asymptotically large distances 
from the wall logarithmically.
The particle number density behaves as
\begin{equation}
n(x) = f_0 \, {\rm e}^{\phi(x)} = 2\pi\ell_{\rm B}\sigma^2 \frac{1}{(1+\widetilde{x})^2} 
\mathop{\sim}_{x\to\infty} \frac{1}{2\pi\ell_{\rm B}} \frac{1}{x^2} .
\label{eq:oneplatedensity}
\end{equation}
The value of the number density at $x=0$, $n(0) = 2\pi\ell_{\rm B}\sigma^2$,
is in agreement with the contact theorem \cite{Henderson78,Henderson79,Choquard80,Carnie81,Totsuji81,Wennerstrom82,Mallarino15}. 
We further see that the large-distance decay of the particle number density
is universal, independent of the surface charge density $\sigma e$:
the only restriction is that $\sigma\ne 0$. This well known but remarkable result 
illustrates in a particular strong form a 
phenomenon of saturation, considered as a hallmark of Poisson-Boltzmann
theory: upon increasing the charge of a field-creating macroion,
one eventually reaches a regime where the electrostatic signature
becomes independent of the macroion 
charge \cite{BoTA02,TeTr03}\footnote{Interestingly, we note that this level of universality still holds, 
at large distances, for arbitrary Coulombic couplings, including thus those that do violate the mean-field/Poisson-Boltzmann 
assumption. We expect physics at large scales to locally fall in the mean-field category, see point 5.3 in \cite{Varenna}.}.
Here, not only is saturation observed at finite $x$ increasing
$\sigma$ (and thus letting $\mu\to 0$), but it is also met --
and this is specific to one dimensional geometry --
at any finite $\sigma$ for $x\to \infty$. In both cases, this
is a signature of efficient screening. We will see below that these
properties are lost for certain classes of polydisperse 
counterionic systems, where screening is impeded by counterions
of a too small valence.

\subsection{Polydisperse case}
We now consider counterions with charges $-q e$, where $q$ is constrained to 
the interval $[0,1]$. 
{The upper bound is arbitrary and rather than some $q_{\text{max}}$, we take it to be unity for the sake of convenience.
We stress here that when results are rescaled with the mean value $\langle q\rangle$, they become 
independent of the choice of $q_{\text{max}}$}. 
The model is defined by a density distribution (per unit surface)
$n(q)$ of particles with the charge $-q e$.
The distribution $n(q)$ might be discrete, i.e. it is a sum of 
$\delta$-functions, or continuous; for the next treatment, we consider that
$n(q)$ is continuous at least close to  $q=0$.
We define the (normalised) moments of the $n(q)$-distribution as follows
\begin{equation} \label{moments}
\langle q^j \rangle \equiv 
\frac{\int_0^1 {\rm d}q\, q^j n(q)}{\int_0^1 {\rm d}q\, n(q)} , \qquad
j=1,2,\ldots .
\end{equation}

Within the PB theory, the density of particles with charge $q$ 
at distance $x$ from the wall, $n(q,x)$, is expressed as
\begin{equation}
n(q,x)\, =\, f(q) \,{\rm e}^{q\phi(x)} ,
\end{equation}
where $f(q)$ is a positive normalisation function; it was equal
to $f_0 \, \delta(q-1)$ in the monodisperse system. 
From this relation, the total particle number density at $x$ is given by
\begin{equation} \label{dens}
n(x) = \int_0^1 {\rm d}q\, n(q,x) = \int_0^1 {\rm d}q\, 
f(q) {\rm e}^{q\phi(x)} .  
\end{equation}
The charge density at $x$ is expressible as
\begin{equation} \label{charge}
\rho(x) = \int_0^1 {\rm d}q\, (-e q) n(q,x) \, = \,
-e \, \int_0^1 {\rm d}q\, q f(q) {\rm e}^{q\phi(x)}
= -e \,\frac{n'(x)}{\phi'(x)} . 
\end{equation}
The number density distribution $n(q)$ is given by
\begin{equation} \label{nfself}
n(q) = \int_0^{\infty} {\rm d}x\, n(q,x) = 
f(q) \int_0^{\infty} {\rm d}x\, {\rm e}^{q\phi(x)} .
\end{equation}
This equation relates the density distribution $n(q)$ and the normalisation
function $f(q)$, provided that the reduced potential $\phi(x)$ is known. 
The overall electroneutrality of the system leads to a constraint for $n(q)$: 
\begin{equation}
\sigma \,=\, \int_0^{\infty} {\rm d}x\, \frac{\rho(x)}{-e} 
\, =\, \int_0^1 {\rm d} q\, q \, n(q) .
\end{equation}
Here, it is worth pointing to a subtlety, that lies in the difference
between $f(q)$ and $n(q)$. In a ``particle'' based model, such as a Monte
Carlo simulation, one chooses the identity of the counterion,
thereby fixing the function $n(q)$. Then, $f(q)$ follows in a non-trivial way,
from measuring the equilibrium density profiles of $q$-species. 
On the other hand, in a ``field'' based formulation such as PB theory,
one needs to know $f(q)$ to be able to write the differential
equation to be solved. Starting from $n(q)$, this requires the knowledge 
of the potential $\phi(x)$, which is precisely the object we are looking
for.
This difficulty is essentially absent in the monodisperse
case; it is the main complication to be addressed when considering
polydisperse mixtures.

Inserting $\rho(x)$ (\ref{charge}) into the Poisson equation, we get 
the polydisperse PB equation 
\begin{equation} \label{PB1}
\frac{{\rm d}^2\phi(x)}{{\rm d}x^2} \, = \, 4\pi \ell_{\rm B} 
\, \int_0^1 {\rm d}q\, q \,f(q) \, {\rm e}^{q\phi(x)} .
\end{equation} 
The gauge (\ref{gauge}) and the BCs (\ref{BCreduced}) remain unchanged, i.e.
\begin{equation}
\phi(0) = 0 , \qquad \phi'(0) = - 4\pi \ell_{\rm B}\sigma , \qquad
\phi'(\infty) = 0 .
\end{equation}
As before, $\phi(x)$ goes to $-\infty$ at asymptotically large $x$.
The problem of the polydisperse PB formulation, alluded to above, is that the available 
information about the charge mixture is encoded in the density distribution 
of the charged particle $n(q)$, and not in the normalisation function $f(q)$.
But the natural (or at least analytically convenient) formulation 
is in fact inverse: with a prescribed 
normalisation function $f(q)$, one should solve the PB equation (respecting 
the corresponding BCs) for the reduced potential and then obtain 
the density distribution $n(q)$ of the charged particles by using 
the $n-f$ relation (\ref{nfself}). We explain in the appendix how this
complication was circumvented for numerical purposes. As far as analytical
results are concerned, the ``implicit'' formulation of
Eq. (\ref{PB1}) is not an issue.

As in the monodisperse case, the PB equation (\ref{PB1})
can be integrated into
\begin{equation} \label{PBintegrated}
\left[ \phi'(x) \right]^2 = 8 \pi \ell_{\rm B} \int_0^1 {\rm d}q\,
f(q) {\rm e}^{q\phi(x)} .
\end{equation}  
The integration constant is again equal to $0$ due to the BCs
$\phi'(x)\to 0$ and ${\rm e}^{\phi(x)}\to 0$ in the limit $x\to\infty$.
The gauge and the BC at $x=0$ imply the constraint
\begin{equation} \label{const}
\int_0^1 {\rm d}q\, f(q) = 2\pi \ell_{\rm B} \sigma^2 ,
\end{equation} 
which is equivalent to the fact that, according to the contact theorem 
\cite{Henderson78,Henderson79,Choquard80,Carnie81,Totsuji81,Wennerstrom82,Mallarino15}, the contact density $n(0)=2\pi\ell_{\rm B}\sigma^2$. 
Equation (\ref{PBintegrated}) can be rewritten as follows
\begin{eqnarray} 
\left[ \phi'(x) \right]^2 & = & 
8 \pi \ell_{\rm B} \int_0^1 {\rm d}q\, f(q) {\rm e}^{-q[-\phi(x)]} \nonumber \\
& = & 8 \pi \ell_{\rm B} \int_0^{-\phi(x)} \frac{{\rm d}p}{[-\phi(x)]}
f\left( \frac{p}{-\phi(x)}\right) {\rm e}^{-p} . \label{rewrite} 
\end{eqnarray}
In the polydisperse case, we define the dimensionless distance as 
\begin{equation} 
\widetilde{x}\equiv \frac{x}{\mu} , \qquad 
\mu=\frac{1}{2\pi\ell_{\rm B} \sigma \langle q\rangle} .
\end{equation}
Note that this definition is consistent with that used in the monodisperse
case (\ref{xtilde}) for which $\langle q\rangle = 1$.
We know that in the limit $x\to\infty$ the function $-\phi(x)\to\infty$.
This means that at asymptotically large distances only the leading
small-$q$ term of the positive distribution function $f(q)$ matters in 
Eq. (\ref{rewrite}). It is also clear that $f(q)\to 0$ for $q\to 0$.
From Eq. (\ref{nfself}) indeed, this is the only way to ensure 
a non-divergent surface density $n(q)$ for $q\to 0$.
Let us then suppose that 
\begin{equation} \label{fqasymp}
f(q) \mathop{\sim}_{q\to 0} 2\pi \ell_{\rm B} \sigma^2 a \, q^{\alpha} ,
\end{equation}  
where $a>0$ and $\alpha\ge 0$ are some dimensionless parameters;
the presence of the prefactor $2\pi \ell_{\rm B} \sigma^2$ 
is motivated by the constraint (\ref{const}).
Considering this small-$q$ behaviour, we show in Appendix \ref{app:asymptotic}
that it is possible to work out the long-distance asymptotic
for all quantities of interest (charge density, ionic density, electrostatic potential),
where novel scaling laws -- explicitly dependent on exponent $\alpha$ -- do emerge.

A meaningful
way to present the results is to introduce $\gamma=(\alpha-3)/2$, which turns out to characterise the 
small-$q$ behaviour of the charge distribution:
\begin{equation}
\frac{n(q)}{\sigma} \mathop{\sim}_{q\to 0} c q^{\gamma} ,
\end{equation}
which defines the parameters $c>0$ and $\gamma>-1$. The results derived in Appendix
\ref{app:asymptotic} then translate into
\begin{equation}
\phi(\widetilde{x}) \mathop{\sim}_{\widetilde{x}\to\infty} 
- \left[ c \frac{2^{2\gamma+5}}{\sqrt{\pi}} \frac{\gamma+3}{\gamma+2}
\Gamma\left( \gamma+\frac{5}{2} \right) \right]^{\frac{1}{\gamma+3}}
(\widetilde{x}/\langle q\rangle)^{\frac{1}{\gamma+3}} , 
\end{equation}
\begin{eqnarray}
n(\widetilde{x})  \mathop{\sim}_{\widetilde{x}\to\infty} & & 2\pi\ell_{\rm B} \sigma^2
\left[ c \frac{2^{\gamma+2}}{\sqrt{\pi}} \frac{1}{(\gamma+2)(\gamma+3)^{\gamma+2}}
\Gamma\left( \gamma+\frac{5}{2} \right) \right]^{\frac{2}{\gamma+3}}
\frac{1}{(\widetilde{x}/\langle q\rangle)^{2\left(\frac{\gamma+2}{\gamma+3}\right)}} , 
\label{eq:densplateasymptotics}
\end{eqnarray}
\begin{equation}
\frac{\rho(\widetilde{x})}{(-e)} \mathop{\sim}_{\widetilde{x}\to\infty} 
2\pi\ell_{\rm B} \sigma^2
\left[ c \frac{2^{\gamma+2}}{\sqrt{\pi}} \frac{(\gamma+2)^{\gamma+2}}{
(\gamma+3)^{2\gamma+5}}\Gamma\left( \gamma+\frac{5}{2} \right) 
\right]^{\frac{1}{\gamma+3}} 
\frac{1}{(\widetilde{x}/\langle q\rangle)^{\frac{2\gamma+5}{\gamma+3}}} .
\label{eq:noneplateasymptotics}
\end{equation}
In particular, when $n(q)$ goes to a nonzero constant $c$ in 
the limit $q\to 0$, which corresponds to $\gamma=0$, we have
\begin{eqnarray}
\phi(\widetilde{x}) \mathop{\sim}_{\widetilde{x}\to\infty} 
- 6^{2/3} c^{1/3} (\widetilde{x}/\langle q\rangle)^{1/3} , \quad
n(\widetilde{x}) \mathop{\sim}_{\widetilde{x}\to\infty} 2\pi\ell_{\rm B} \sigma^2
\frac{1}{6^{2/3}} c^{2/3} \frac{1}{(\widetilde{x}/\langle q\rangle)^{4/3}} ,
\nonumber \\
\frac{\rho(\widetilde{x})}{(-e)} \mathop{\sim}_{\widetilde{x}\to\infty} 
2\pi\ell_{\rm B} \sigma^2 \left( \frac{2}{9} \right)^{2/3} c^{1/3} 
\frac{1}{(\widetilde{x}/\langle q\rangle)^{5/3}} .
\end{eqnarray}

As we have seen, the non-universal large-distance behaviour of the quantities
like the reduced potential and particle/charge densities for the charge 
mixtures within the PB theory can be related to the small-$q$ behaviour of 
the density distribution $n(q)$.
If there is e.g. a gap in $q$ and the function $f(q)$, or equivalently
$n(q)$, is zero up to some positive threshold $q_{\min}$, the integral in 
(\ref{PBintegrated}) is dominated by $\exp[q_{\min}\phi(x)]$ at large 
distances from the wall, and we basically recover the monodisperse relation 
of type (\ref{intPB}). It is always the population with smallest valence
which sets the large distance asymptotic, and non-trivial effects 
emerge when this population has a vanishing charge ($q_{\text{min}}=0$).
A similar remark holds for the two-plate problem to be discussed below.
Yet, even a discrete charge distribution may exhibit,
transiently,
the power-laws brought to the fore here, see section \ref{ssec:transient} below.

\subsection{Numerical PB results for a simple polydisperse model}

As emphasised above, a physical 
problem is posed specifying the distribution $n(q)$, rather than the normalisation function $f(q)$,
which is unknown without having solved the PB equation, the formulation of which requires
the knowledge of $f(q)$. This question will be addressed in the remainder (see the appendix),
but to circumvent this complication and test the premises of our analytical approach, we have 
chosen the specific form
\begin{equation} \label{deffq}
f(q) = 2\pi \ell_{\rm B} \sigma^2 a q^{\alpha} \qquad q\in [0,1] ,
\end{equation}
with integer $\alpha=2,3,4,\ldots$.
This function represents an extension of the small-$q$ asymptotic
(\ref{fqasymp}) to the whole $q$-interval.
The constraint (\ref{const}) fixes the prefactor $a$ to
\begin{equation}
a = \alpha+1 .
\end{equation}
The PB equation reads as
\begin{equation}
\phi'(\widetilde{x}) \langle q\rangle = - 2 \sqrt{\alpha+1} 
\left[ \int_0^1 {\rm d}q\, q^{\alpha} {\rm e}^{q\phi(\tilde{x})} \right]^{1/2} .
\end{equation}
The advantage of the chosen model is that the function inside the integral on 
the rhs is explicitly integrable:
\begin{eqnarray}
\int_0^1 {\rm d}q\, q^{\alpha} {\rm e}^{q\phi} & = & 
\frac{{\rm e}^{\phi}(\phi^2-2\phi+2)-2}{\phi^3} \qquad \mbox{for $\alpha=2$,}
\nonumber \\ & = & 
\frac{{\rm e}^{\phi}(\phi^3-3\phi^2+6\phi-6)+6}{\phi^4} \qquad 
\mbox{for $\alpha=3$,}
\nonumber \\ & = & 
\frac{{\rm e}^{\phi}(\phi^4-4\phi^3+12\phi^2-24\phi+24)-24}{\phi^5} \qquad 
\mbox{for $\alpha=4$,}
\end{eqnarray}
etc. 
This allows us to solve numerically the PB equation in a particularly straightforward
manner.

\begin{figure}
\begin{center}
\includegraphics[width=0.5\textwidth,clip]{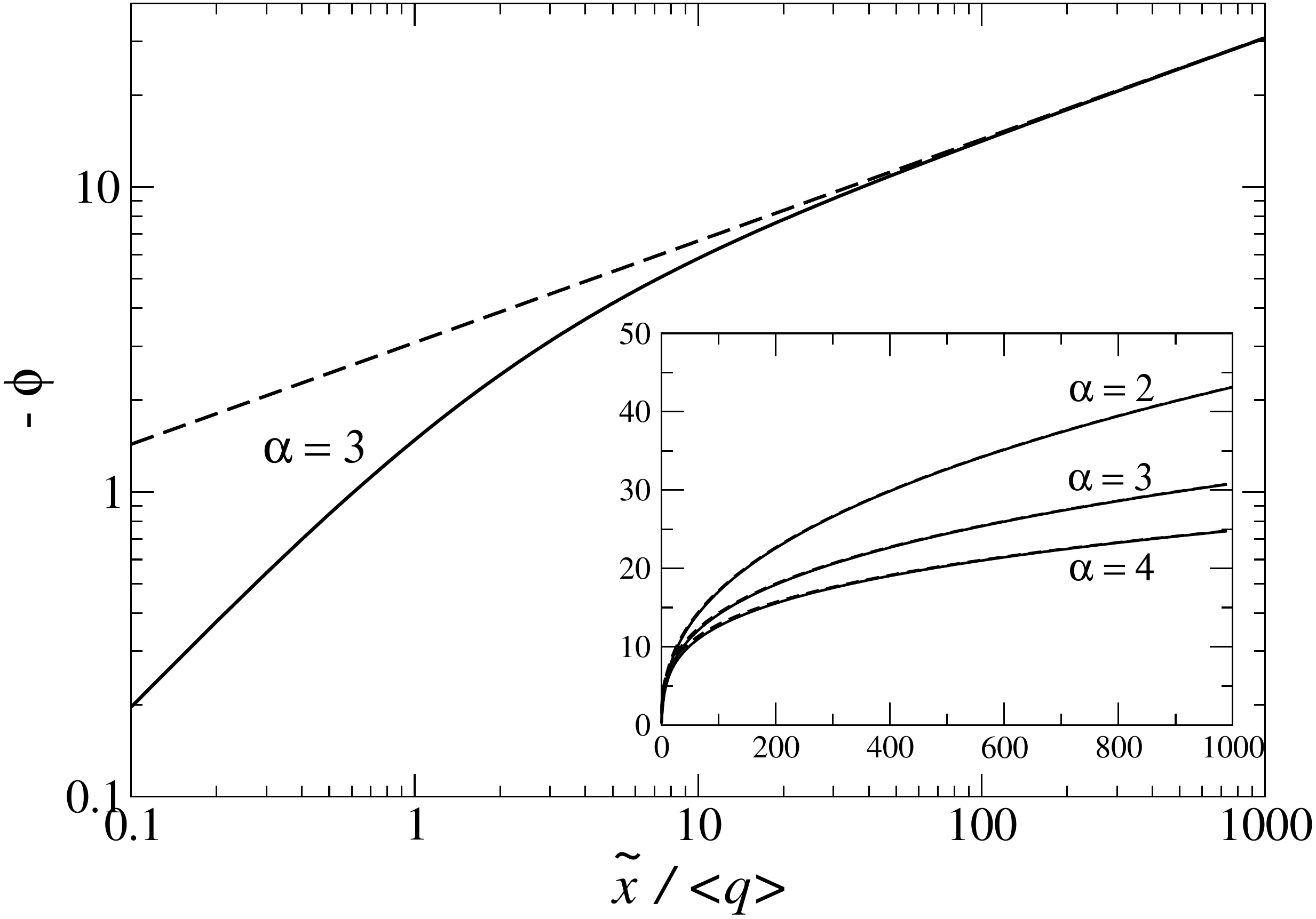}
\caption{Log-log plot of the (minus) potential $-\phi$ vs. the reduced 
distance $\widetilde{x}/\langle q\rangle$ for $\alpha=3$. 
The solid curve corresponds to the numerical treatment of the PB equation
with the normalisation function $f(q)$ defined by Eq. (\ref{deffq}),
the dashed line corresponds to the asymptotic formula (\ref{phiasym}).
Inset: same results on a linear scale, for $\alpha=2, 3$ and 4. The dashed and
continuous lines are essentially superimposed.}
\label{fig:pot} 
\end{center}
\end{figure}

For $\alpha=2,3,4$, the numerical results for the electric 
potential $\phi$ versus distance are presented by the solid curve in Fig. \ref{fig:pot}.
For comparison, the analytically obtained asymptotic formula for the potential 
(\ref{phiasym}) are represented by the dashed lines.
We see that the asymptotic regime is already reached at 
$\widetilde{x}/\langle q\rangle\sim 100$.

\begin{figure}[htb]
\begin{center}
\includegraphics[width=0.5\textwidth,clip]{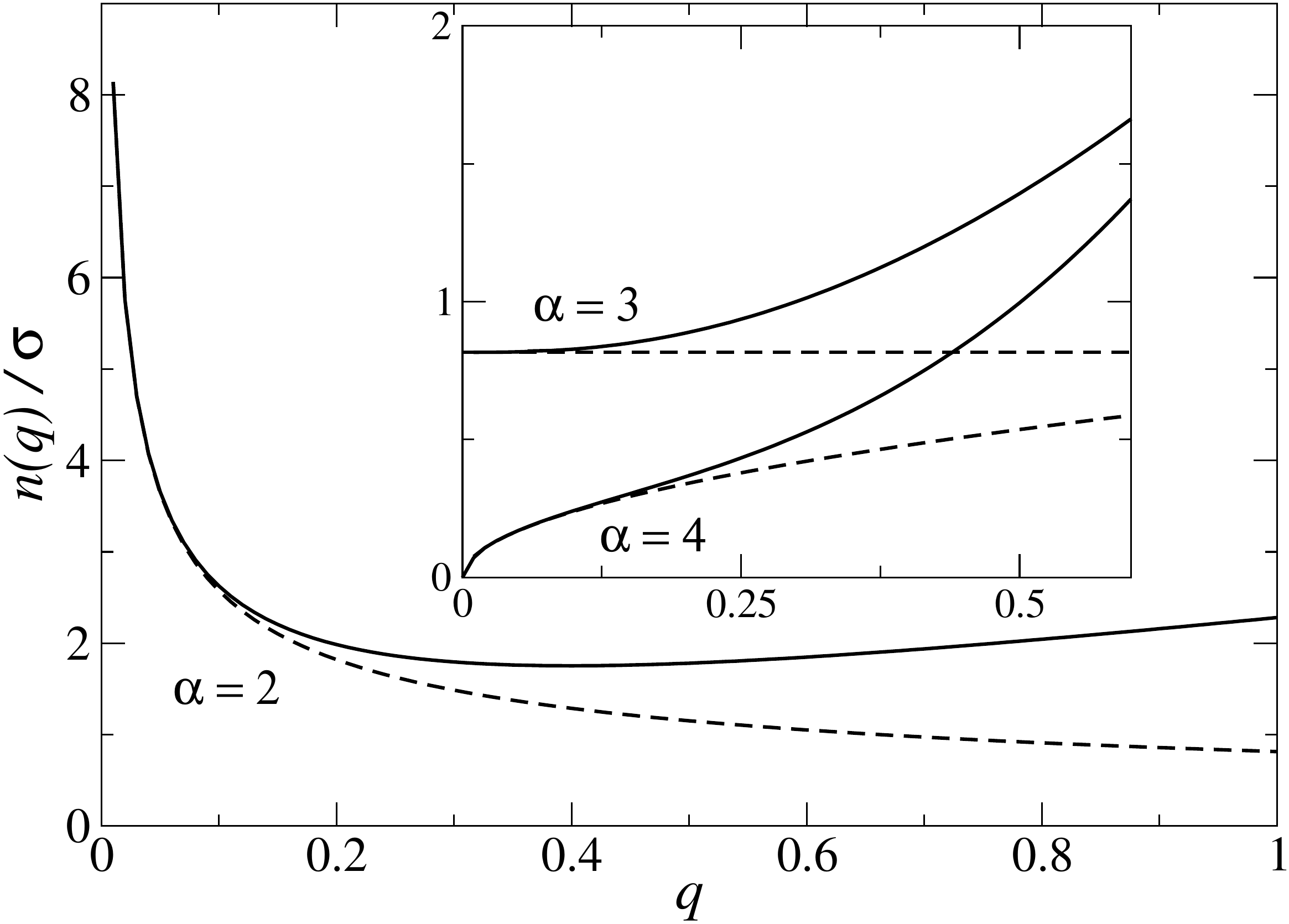}
\caption{Plot of the particle density  distribution $n(q)/\sigma$ vs. 
valence $q$ for $\alpha=2, 3, 4$. For each $\alpha$,
the solid curve corresponds to the numerical solution of the PB equation
with the normalisation function $f(q)$ defined by Eq. (\ref{deffq}), while
the dashed curve is for the asymptotic formula (\ref{finalrel}).}
\label{fig:density} 
\end{center}
\end{figure}

Having at our disposal the function $\phi(\widetilde{x})$, we calculate
the  particle density distribution $n(q)/\sigma$, which corresponds to 
our model (\ref{deffq}), by using the formula (\ref{nq}).
For $\alpha=2,3,4$, the numerical solution of the PB equation is 
represented by the solid curve in Fig. \ref{fig:density}\footnote{
In order to obtain adequate results
in the small $q$ region, the integral over $\widetilde{x}$ in (\ref{nq}) 
has to be computed on a very large interval, ranging from 10 to 100 millions 
of length units.}.
The dashed curves appearing in this figure correspond to the asymptotic 
$q\to 0$ formula (\ref{finalrel}).
As $q\to 0$, the density distribution diverges for $\alpha=2$, attains
a finite value for $\alpha=3$ and goes to 0 for $\alpha=4$. 
We see that the agreement of the numerical and analytical calculations
in the small-$q$ region is good.
This confirms our previous assumption that the small-$q$ asymptotics of 
the functions $n(q)$ and $f(q)$ are related by Eq. (\ref{smallq}) with 
the asymptotic large-$\widetilde{x}$ potential (\ref{phiasym}) inserted.

\section{Symmetric two-wall geometry}
\label{sec:twowalls}
Now we consider a symmetric pair of parallel hard walls of dielectric constant 
$\epsilon$ at distance $d$.
Each of the surfaces at $x=-d/2$ and $x=d/2$ carries a constant surface 
charge density $\sigma e$ with $\sigma>0$.
The charged particles, confined to the slit $-d/2<x<d/2$, are immersed in 
a medium of the same dielectric constant as the walls, i.e.~$\epsilon$.
The electroneutrality condition reads as
\begin{equation} \label{neutrality2}
2 \sigma e + \int_{-d/2}^{d/2} {\rm d}x\, \rho(x) = 0 .
\end{equation} 
Integrating the Poisson equation (\ref{Poisson}) from $-d/2$ to $d/2$,
the condition (\ref{neutrality2}) is consistent with the couple of BCs
\begin{equation} \label{BC2}
\psi'\left(-\frac{d}{2}\right) = - \frac{4\pi\sigma e}{\epsilon} , \qquad 
\psi'\left(\frac{d}{2}\right) = \frac{4\pi\sigma e}{\epsilon} .
\end{equation} 
The problem is symmetric with respect to the sign reversal of 
the $x$-coordinate, i.e. $\psi(x)=\psi(-x)$, $n(x)=n(-x)$, $\rho(x)=\rho(-x)$.
Consequently, $\psi'(x)=-\psi'(-x)$, $n'(x)=-n'(-x)$, $\rho'(x)=-\rho'(-x)$, 
so the derivatives of these quantities vanish at $x=0$.
In particular,
\begin{equation} \label{BC3}
\psi'(0) = 0 .
\end{equation}
This BC formally corresponds to having an uncharged hard wall at $x=0$.

For the subsequent analysis, it turns
out that two equivalent formulations are of particular interest.
They correspond to each other through  a $x\to d/2-x$ transformation,
with a additional shift of potential to enforce the chosen gauge.
\begin{itemize}
\item
(i) In analogy with the one-plate problem, we shift the reference 
to the surface of one of the walls, say the one at $x=-d/2$, and consider
the asymmetric configuration of one charged hard wall at $x=0$ with 
uniform surface charge density $\sigma e$, and one 
uncharged ($\sigma'=0$) plain hard wall at $x= d/2$.
The gauge condition and the corresponding BCs for the reduced potential read as
\begin{equation} \label{BCi} 
\phi(0) = 0 , \qquad \phi'(0) = - 4\pi \ell_{\rm B} \sigma , \qquad 
\phi'(d/2) = 0 .
\end{equation} 
Both $\phi(x)$ and $\phi'(x)$ are negative (or 0) in the whole
interval $[0,d/2]$.
In the monodisperse case, we set $n(x) = f_0 \, {\rm e}^{\phi(x)}$ for the
particle density and the resulting PB equation can be integrated into
\begin{equation} \label{intPBmono1}
\left[ \phi'(x) \right]^2 = 8\pi\ell_{\rm B} f_0 \, \left[ {\rm e}^{\phi(x)} -1 \right]
+ (4\pi\ell_{\rm B}\sigma)^2 .
\end{equation}
Since the confining surfaces are planar, the particle densities at 
contact with the surfaces obey the contact theorem \cite{Henderson78,Henderson79,Choquard80,Carnie81,Totsuji81,Wennerstrom82,Mallarino15}
\begin{equation} \label{ct2}
n(0) = 2\pi \ell_{\rm B} \sigma^2 + \beta P , \qquad
n\left(\frac{d}{2}\right) = \beta P ,
\end{equation} 
where $P$ is the pressure.
Equivalently,
\begin{equation} \label{pressuremonoi}
\beta P = f_0 \, - 2\pi \ell_{\rm B} \sigma^2 , \qquad
\beta P = f_0 \, {\rm e}^{\phi(d/2)} .
\end{equation} 
In the polydisperse case with 
$n(x) = \int_0^1 {\rm d}q\, f(q) {\rm e}^{q\phi(x)}$,
the PB equation can be integrated into
\begin{equation} \label{intPBpoly1}
\left[ \phi'(x) \right]^2 = 8\pi\ell_{\rm B}  \int_0^1 {\rm d}q\, 
f(q) \left[ {\rm e}^{q\phi(x)} -1 \right] + (4\pi\ell_{\rm B}\sigma)^2 .
\end{equation}
The pressure can be written
\begin{equation} \label{pressurepolyi}
\beta P \, = \, \int_0^1 {\rm d}q\, f(q) - 2\pi \ell_{\rm B} \sigma^2  \, = \, \int_0^1 {\rm d}q\, f(q) {\rm e}^{q \phi(d/2)} .
\end{equation} 

\item
(ii) Next, we shift the reference to the midpoint between the walls and consider
the configuration of one uncharged ($\sigma'=0$) hard wall at
$x= 0$ and the charged wall at $x=d/2$ with the (surface charge 
density $\sigma e$).
The gauge condition and the corresponding BCs read as
\begin{equation} \label{BCii}
\phi(0) = 0 , \qquad \phi'(0) = 0 , \qquad 
\phi'(d/2) = 4\pi \ell_{\rm B} \sigma .
\end{equation} 
Both $\phi(x)$ and $\phi'(x)$ are positive (or 0) in the 
interval $[0,d/2]$.
In the monodisperse case, the PB equation is integrated into
\begin{equation} \label{intPBmono2}
\left[ \phi'(x) \right]^2 = 
8\pi\ell_{\rm B} f_0 \, \left[ {\rm e}^{\phi(x)} -1 \right] .
\end{equation}
The pressure is expressible as
\begin{equation} \label{pressuremonoii}
\beta P = f_0 \,=\,
f_0 \, {\rm e}^{\phi(d/2)} - 2\pi \ell_{\rm B} \sigma^2 .
\end{equation} 
On the other hand, the polydisperse PB equation can be integrated into
\begin{equation} \label{intPBpoly2}
\left[ \phi'(x) \right]^2 = 8\pi\ell_{\rm B}  \int_0^1 {\rm d}q\, 
f(q) \left[ {\rm e}^{q\phi(x)} -1 \right] ,
\end{equation}
and the pressure is given by
\begin{equation} \label{pressurepolyii}
\beta P = \int_0^1 {\rm d}q\, f(q)  \, = \,
\int_0^1 {\rm d}q\, f(q) {\rm e}^{q \phi(d/2)} 
- 2\pi \ell_{\rm B} \sigma^2 .
\end{equation} 
\end{itemize}
Note that the explicit form of the normalisation function $f(q)$
depends on the formulation, while $\beta P$ does not.

\subsection{Monodisperse case}
In the monodisperse case with particles of charge $-e$, the solution 
is well known. For completeness, it is reminded here. We use formulation (ii) with the gauge and BCs of type
(\ref{BCii}).
The PB equation (\ref{intPBmono2}), written as
\begin{equation}
\phi'(x) = 2 K \sqrt{{\rm e}^{\phi(x)}-1} , \qquad 
K = \sqrt{2\pi\ell_{\rm B} f_0} ,
\end{equation}
has the explicit solution
\begin{equation}
\phi(x) = - 2 \ln \cos(Kx).
\end{equation}
The BC at $x=d/2$ implies the transcendental equation for the screening parameter $K$:
\begin{equation} \label{Kdeq}
K d \tan\left( K \frac{d}{2} \right) = 2\pi\ell_{\rm B}\sigma d
\equiv \widetilde{d} .
\end{equation}

In the limit $\widetilde{d}\to 0$, $K d$ is small and one can expand
Eq. (\ref{Kdeq}) in powers of $K d$ to obtain the small-distance behaviour
of the pressure $\beta P = f_0$,
\begin{equation} \label{Psmall}
\tilde{P} \equiv \frac{\beta P}{2\pi \ell_{\rm B}\sigma^2} = \frac{2}{\widetilde{d}} 
- \frac{1}{3} + \frac{2}{45} \widetilde{d} + \cdots .
\end{equation} 

In the large-distance limit $\widetilde{d}\to\infty$, we have $K d \to \pi$. 
The pressure
\begin{equation} \label{Plarge}
\beta P ~ \mathop{\sim}_{d\to\infty} ~ \frac{\pi}{2\ell_{\rm B}} \, \frac{1}{d^2}
\end{equation}
is then independent of the surface charge density $\sigma e$. 
It is interesting to compare this result to the  one-plate density,
as given by Eq. (\ref{eq:oneplatedensity}). In the present
salt-free problem, the superposition of the two 1-plate densities
$n_1(x)+n_1(d-x)$ is never a good approximation to
the complete two-plates profiles. Yet, following that incorrect route
to compute the pressure, we get the correct scaling in $1/d^2$ for the 
pressure, with a prefactor $4/(\pi\ell_{\rm B})$ instead of $\pi/(2\ell_{\rm B})$
as given by Eq. (\ref{Plarge}). The ratio of both is thus ${\cal R} = \pi^2/8\simeq 1.23$,
and can be seen as a quantitative measure of (pressure enhancing) non-linear effects. It will be
seen that this ratio is significantly larger in the polydisperse case.

Both short- and large-distance expansions can be derived systematically 
in alternative ways, without solving explicitly the model.
Since these alternative techniques are important for the polydisperse
case, we shall review them in the following. 

\subsubsection{Short-distance expansion}
We still use formulation (ii) with the gauge and BCs of type
(\ref{BCii}) and the PB equation (\ref{intPBmono2}).
Since the electric potential measured from the midpoint $x=0$ has 
the symmetry $\phi(x) = \phi(-x)$, its small-$x$ expansion reads as
\begin{equation}
\phi(x) = a_1 x^2 + a_2 x^4 + \cdots .
\end{equation}
Inserting this expansion into the PB equation (\ref{intPBmono2}),
the expansion coefficients are given by
\begin{equation}
a_1 = 2 (\pi\ell_{\rm B}f_0) , \qquad
a_2 = \frac{2}{3} (\pi\ell_{\rm B}f_0)^2 ,
\end{equation}
etc.
The normalisation condition
\begin{equation}
2\sigma = \int_{-d/2}^{d/2} {\rm d}x\, n(x) 
= 2 f_0 \, \int_0^{d/2} {\rm d}x\, {\rm e}^{\phi(x)}
\end{equation}
together with the small-$d$ expansion of the integral
\begin{equation}
\int_0^{d/2} {\rm d}x\, {\rm e}^{a_1 x^2+a_2 x^4+\cdots} 
= \frac{d}{2} + \frac{1}{12} (\pi\ell_{\rm B}f_0) d^3 
+ \frac{1}{60} (\pi\ell_{\rm B}f_0)^2 d^5 + \cdots
\end{equation}
can be used to derive a small-$d$ expansion for $f_0$:
\begin{equation}
f_0 = \frac{2\sigma}{d} - \frac{2}{3} \pi \ell_{\rm B} \sigma^2
+ \frac{8}{45} (\pi \ell_{\rm B})^2 \sigma^3 d + \cdots .
\end{equation}
With regard to the relation $\beta P = f_0$, we end up with 
the short-distance expansion (\ref{Psmall}).

\subsubsection{Large-distance expansion}
With the same gauge and BCs as in the previous part, the PB equation 
(\ref{intPBmono2}) can be re-expressed via the separation of variables as
\begin{equation}
\frac{{\rm d}\phi}{\sqrt{{\rm e}^{\phi}-1}} = 
\sqrt{8\pi\ell_{\rm B}f_0} \, {\rm d}x ,
\end{equation}
which implies
\begin{equation}
\int_0^{\phi(d/2)} \frac{{\rm d}\phi}{\sqrt{{\rm e}^{\phi}-1}} = 
\sqrt{8\pi\ell_{\rm B}f_0} \, \frac{d}{2} .
\end{equation}
In the limit $d\to\infty$ we have $\phi(d/2)\to\infty$ and
the integral on the lhs equals to $\pi$.
This leads to $f_0 = \pi/(2\ell_{\rm B}d^2)$ which is equivalent to 
the anticipated result (\ref{Plarge}).
Note that this approach does not need the explicit PB solution,
which is an interesting feature.

\subsection{Polydisperse case}
With the valence density distribution $n(q)$, 
the definition of the moments (\ref{moments}) and of the dimensionless
distance $\widetilde{x} = 2\pi \ell_{\rm B} \sigma \langle q\rangle x$ remain
unchanged.
Electro-neutrality reads
\begin{equation} \label{neutr2}
\int_0^1 {\rm d}q\, q \, n(q) = 2 \sigma .
\end{equation} 
The normalisation function $f(q)$ in the PB equation is related to 
the number density distribution of charges $n(q)$ via
\begin{equation} \label{rel}
n(q) = f(q) \int_{-d/2}^{d/2} {\rm d}x\, {\rm e}^{q\phi(x)} 
= 2 f(q) \int_0^{d/2} {\rm d}x\, {\rm e}^{q\phi(x)} ,
\end{equation}
where we took into account the reflection symmetry of the potential
with respect to the midpoint between the walls, $\phi(x) = \phi(-x)$.

\subsubsection{Short-distance expansion}
As before, we consider the formulation (ii) with the gauge and BCs of type
(\ref{BCii}), the PB equation (\ref{intPBpoly2}) and the pressure
(\ref{pressurepolyii}).
Around $x=0$, the reduced potential is searched in the form
\begin{equation}
\phi(x) = a_1 x^2 + a_2 x^4 + \cdots .
\end{equation}
Inserting this expansion into the PB equation (\ref{intPBpoly2}) and comparing 
the $x$-powers on both sides, the expansion coefficients are given by
\begin{eqnarray}
a_1 & = & 2 \pi\ell_{\rm B} \int_0^1 {\rm d}q\, q f(q) , \label{a1} \\
a_2 & = & \frac{2}{3} (\pi\ell_{\rm B})^2 
\left[ \int_0^1 {\rm d}q\, q f(q) \right]
\left[ \int_0^1 {\rm d}q\, q^2 f(q) \right] , \label{a2} 
\end{eqnarray}
and so on.
The relation (\ref{rel}) implies 
\begin{equation} \label{nf}
n(q) = f(q) \left[ d + \frac{1}{12} a_1 q d^3 
+ \frac{1}{80} \left( a_2 q + \frac{1}{2} a_1^2 q^2 \right) d^5
+ \cdots \right] .
\end{equation}

In the lowest small-$d$ order, it follows from (\ref{nf}) that $f(q)$ 
and $n(q)$ are related via
\begin{equation} \label{fn1}
f(q) = \frac{n(q)}{d} .
\end{equation}
The corresponding pressure reads
\begin{equation}
\beta P = \int_0^1 {\rm d}q\, f(q) = \frac{1}{d} \int_0^1 {\rm d}q\, n(q) . 
\end{equation}
To leading order, both $f(q)$ and $\beta P$ are proportional to $1/d$.
This is nothing but the ideal gas law, valid under extreme confinement,
where the entropy cost for squeezing the ions in a narrow slit 
overweights Coulombic contributions.

In the next order, we find from (\ref{nf}) that
\begin{equation} 
f(q) = \frac{n(q)}{d} - \frac{a_1}{12} \,q\, n(q)\, d .
\end{equation}
The coefficient $a_1$ is expressed in terms of the function $f(q)$ in
Eq. (\ref{a1}).
To compute it, it is sufficient
to take the $f-n$ relation (\ref{fn1}) from the preceding order, i.e.
\begin{equation}
a_1 = 2 \pi\ell_{\rm B} \int_0^1 {\rm d}q\, q \frac{n(q)}{d} .
\end{equation} 
Thus we get, in the $d^0$ order
\begin{equation} \label{fn2}
f(q) = \frac{n(q)}{d} - \frac{\pi \ell_{\rm B}}{6} q n(q) 
\left[ \int_0^1 {\rm d}q\, q n(q) \right] 
\end{equation}
and
\begin{equation} 
\beta P = \frac{1}{d} \int_0^1 {\rm d}q\, n(q)
 - \frac{\pi \ell_{\rm B}}{6} \left[ \int_0^1 {\rm d}q\, q n(q) \right]^2 . 
\end{equation}

In the next order, the relation (\ref{nf}) implies that
\begin{equation} 
f(q) = \frac{n(q)}{d} - \frac{a_1}{12} q n(q) d 
+ \left( \frac{a_1^2}{1440} q^2 - \frac{a_2}{80} q \right) n(q) d^3 .
\end{equation}
The coefficient $a_1$ in Eq. (\ref{a1}) is calculated using the function $f(q)$
from the preceding $f-n$ relation (\ref{fn2}), 
\begin{equation}
a_1 = \frac{2 \pi\ell_{\rm B}}{d} \int_0^1 {\rm d}q\, q n(q)
- \frac{1}{3} (\pi\ell_{\rm B})^2 \left[ \int_0^1 {\rm d}q\, q n(q) \right]
\left[ \int_0^1 {\rm d}q\, q^2 n(q) \right] ,
\end{equation} 
while to obtain the coefficient $a_2$ (\ref{a2}) at the correct order it is 
sufficient to take the function $f(q)$ from the $f-n$ relation (\ref{fn1}), 
\begin{equation}
a_2 = \frac{2}{3} \frac{(\pi\ell_{\rm B})^2}{d^2} 
\left[ \int_0^1 {\rm d}q\, q n(q) \right]
\left[ \int_0^1 {\rm d}q\, q^2 n(q) \right] .
\end{equation}
In the $d^1$ order, we arrive at
\begin{eqnarray} \label{fn3}
f(q) & = & \frac{n(q)}{d} - \frac{\pi \ell_{\rm B}}{6} 
\left[ \int_0^1 {\rm d}q\, q n(q) \right] q n(q) 
+ \frac{7}{360} (\pi \ell_{\rm B})^2 d \left[ \int_0^1 {\rm d}q\, q n(q) \right] 
\nonumber \\ & & \times \left[ \int_0^1 {\rm d}q\, q^2 n(q) \right] q n(q) 
+ \frac{1}{360} (\pi \ell_{\rm B})^2 d 
\left[ \int_0^1 {\rm d}q\, q n(q) \right]^2 q^2 n(q) 
\end{eqnarray}
and
\begin{eqnarray} 
\beta P & = & \frac{1}{d} \int_0^1 {\rm d}q\, n(q)
- \frac{\pi \ell_{\rm B}}{6} \left[ \int_0^1 {\rm d}q\, q n(q) \right]^2
\nonumber \\ & & 
+ \frac{1}{45} (\pi \ell_{\rm B})^2 d \left[ \int_0^1 {\rm d}q\, q n(q) \right]^2 
\left[ \int_0^1 {\rm d}q\, q^2 n(q) \right] . 
\end{eqnarray}
Taking into account the electroneutrality condition (\ref{neutr2}), the pressure 
can be rewritten in terms of the moments (\ref{moments}) and the dimensionless distance 
$\widetilde{d} = 2\pi \ell_{\rm B} \sigma \langle q\rangle d$
as
\begin{equation} \label{Psmall2} 
\tilde{P} \equiv \frac{\beta P}{2\pi\ell_{\rm B}\sigma^2} = \frac{2}{\widetilde{d}} - \frac{1}{3}
+ \frac{2}{45} \widetilde{d} \frac{\langle q^2\rangle}{\langle q\rangle^2} . 
\end{equation}
Notice that the first two terms of this small-$\widetilde{d}$ expansion
do not depend on the number distribution $n(q)$.
For the monodisperse case with the distribution $n(q)=2\sigma \delta(q-1)$ 
and the moments $\langle q^j\rangle = 1$ for all $j=1,2,\ldots$ we recover
the previous result (\ref{Psmall}).
For the uniform distribution $n(q)=4\sigma$ with the moments 
$\langle q^j\rangle = 1/(j+1)$ $(j=1,2,\ldots)$, the third term on
the rhs of (\ref{Psmall2}) is modified by the factor 
$\langle q^2\rangle/\langle q\rangle^2 = 4/3$. 

The method presented in this part works not only for continuous distributions
$n(q)$, but also for discrete distributions like 
$n(q)=\sum_{\alpha=1}^{Q} n_{\alpha} \delta(q-q_{\alpha})$ 
involving counterions of the same sign. 

\subsubsection{Large-distance expansion}
For simplicity, let us restrict ourselves to the interesting {case} 
having uniform density distribution ($n(q)=4\sigma$ with $q\in [0,1]$ so that
$\langle q\rangle = 1/2$), {both because of its simplicity and because it is, loosely speaking, ``maximally''
distinct from the discrete cases studied previously}. 

We switch to the formulation (i) with the gauge and BCs of type (\ref{BCi}). 
The PB equation (\ref{intPBpoly1}) is rewritten as
\begin{equation} \label{PBnew}
\left[ \phi'(x) \right]^2 = 8\pi\ell_{\rm B} \left[
\int_0^1 {\rm d}q\, f(q) {\rm e}^{q\phi(x)} - \beta P \right] 
\end{equation} 
and the pressure is given by (\ref{pressurepolyi}). Other choices of distribution with $q_{\text{max}}\neq 1$ can be recast into
Eq.~(\ref{PBnew}) with $\ell_B$ and $q$ replaced by $\ell_B q^2_{\text{max}}$ and $q/q_{\text{max}}$ respectively.   
We keep in mind that both $\phi(x)$ and $\phi'(x)$ are negative or equal to
0 on the whole interval $[0,d/2]$.

We assume that for large distance $d$ the potential behaves like 
in the one-wall case (\ref{asymppot}) with the same exponent $\beta=1/3$,
\begin{equation} \label{potas}
\phi(\widetilde{x}) \sim - b 
\left( \frac{\widetilde{x}}{\langle q\rangle} \right)^{1/3} .
\end{equation}
Here, the prefactor $b$, which differs from its one-wall counterpart, 
is as-yet undetermined and the dimensionless distance is
$\widetilde{x} = 2\pi\ell_{\rm B}\sigma \langle q\rangle x$.
As in the one-wall problem, we expect that for large $d$,
the relation (\ref{rel}) between $n(q)$ and $f(q)$ is determined for
small $q$ by the asymptotic form of $\phi(x)$.
Inserting (\ref{potas}) into (\ref{rel}) results in
\begin{equation} \label{fexp}
\frac{f(q)}{2\pi\ell_{\rm B}\sigma^2} \,= \, \frac{4\langle q\rangle}{\widetilde{d}} \,
h(u) , \qquad u = b q \left( \frac{\widetilde{d}}{2\langle q\rangle} \right)^{1/3} ,
\end{equation}
where 
\begin{equation}
h(u) \, = \,  \frac{u^3/3}{2-{\rm e}^{-u}(2+2u+u^2)} 
\, \mathop{\sim}_{u\to 0}  \, 1 + \frac{3}{4} u + \frac{21}{80} u^2 + \cdots
\end{equation}
On the other hand, when $\widetilde d$ (and thus $u$) is large,
we get
\begin{equation}
h(u)  \mathop{\sim}_{u\to \infty} u^3/6 \quad \hbox{and} \quad
\frac{f(q)}{2\pi\ell_{\rm B}\sigma^2}  \mathop{\sim}_{u\to \infty} 
b^3 q^3/3,
\label{asympf}
\end{equation}
Note that for large but finite $\widetilde{d}$, the value of 
\begin{equation}
\frac{f(0)}{2\pi\ell_{\rm B}\sigma^2} = \frac{4\langle q\rangle}{\widetilde{d}}
= \frac{2}{\widetilde{d}}
\end{equation}
does not vanish. This is at variance with the one-wall case formulated in an unconstrained 
half-space, where $f(q) \propto q^3$ for small $q$ (we are indeed addressing the situation
where $n(q)$ goes to a constant for $q\to 0$, so that $\gamma=0$ and $\alpha=3$).
The pressure is given by
\begin{eqnarray} 
\tilde{P} \equiv \frac{\beta P}{2\pi\ell_{\rm B}\sigma^2} & = & \displaystyle \int_0^1 {\rm d}q\, 
\frac{f(q)}{2\pi\ell_{\rm B}\sigma^2} \exp\left[- b q 
\left( \frac{\widetilde{d}}{2\langle q\rangle} \right)^{1/3} \right] \nonumber \\
& \displaystyle \mathop{\sim}_{\tilde{d}\to\infty} &
\frac{2}{b} \left( \frac{2\langle q\rangle}{\widetilde{d}} \right)^{4/3}
\int_0^{\infty} {\rm d}u\, {\rm e}^{-u} h(u) . \label{pressurenew}
\end{eqnarray}
At large distances, $d$ appears under the combination 
$\widetilde d / \langle q \rangle = 2 \pi \ell_B \sigma d$, which 
is independent of the valence distribution, and in particular independent
of $\langle q \rangle$. This simply stems from the fact at large-$d$,
screening is mediated by those ions of smallest valence, irrespective of the details
of the complete distribution. We have already met that statement above.
What makes the present situation of interest is that these ions have a 
vanishing valence. Should this not be the case, one would recover the
monodisperse phenomenology, with as asymptotic decay of pressure in 
$1/d^2$.

To obtain the prefactor $b$, Eq. (\ref{PBnew}) tells us that
for large $\widetilde{d}$ we have
\begin{equation}
\int_0^{-b\left( \frac{\tilde{d}}{2\langle q\rangle} \right)^{1/3}}
\frac{{\rm d}\phi}{\sqrt{\int_0^1 {\rm d}q\, 
\frac{f(q)}{2\pi\ell_{\rm B}\sigma^2}{\rm e}^{q\phi} -
\frac{\beta P}{2\pi\ell_{\rm B}\sigma^2}}} = 
- \frac{\widetilde{d}}{\langle q\rangle} .
\end{equation}
With the aid of the substitution 
\begin{equation}
\phi = -b \left( \frac{\widetilde{d}}{2\langle q\rangle} \right)^{1/3} \varphi 
\end{equation}
the powers of $\widetilde{d}$ correctly cancel on both sides of this equality,
confirming the adequacy of the assumption (\ref{potas}), and we arrive 
at the equation
\begin{equation}
\int_0^1 \frac{{\rm d}\varphi}{\int_0^{\infty} {\rm d}u\, 
( {\rm e}^{-u\varphi} - {\rm e}^{-u}) h(u)} = \left( \frac{2}{b} \right)^{3/2} .
\end{equation}
It implies that $b=3.18623$.
Considering this value in (\ref{pressurenew}) leads to 
the large-distance asymptotic
\begin{equation}
\label{analypPB}
\tilde{P} \equiv \frac{\beta P}{2\pi\ell_{\rm B}\sigma^2} \mathop{\sim}_{\widetilde{d}\to\infty}
\frac{1.766}{(\widetilde{d}\, )^{4/3}} .
\end{equation}
The corresponding exponent, $4/3$, is significantly smaller than 
that holding in the monodisperse case (where $P \propto d^{-2}$)
as yet another signature of less efficient screening, with therefore
an enhanced inter-plate repulsion at large distances.

\subsection{Comparison to numerical results}

We now test our analytical predictions. Once the polydispersity 
function $n(q)$ has been chosen, we a) solve iteratively the
PB equation and b) perform Monte Carlo simulations (MC) at a small
enough Coulombic coupling {($\Xi = 2\pi \langle q\rangle^3 \ell_B^2\sigma<1$)} \cite{NJMN05,SaTr11,Rque03}, which should enforce
the validity of the mean-field PB approach. {For a realistic system of polyvalent ions, 
this constraint can be met if
the surface charge density or/and Bjerrum length is sufficiently small, so that each species-defined coupling parameter, by using
that species valency rather the average one when defining the coupling parameters, is smaller
than one.} Both treatments
are summarised in the appendix. They are very different, one 
consisting in solving an (implicit) differential equation, and
the other one being particle based, with an exact treatment 
of Coulomb forces between each pair of charged bodies (wall-wall,
ion-wall and ion-ion).

\begin{figure}
\begin{center}
\includegraphics[width=0.5\textwidth,clip]{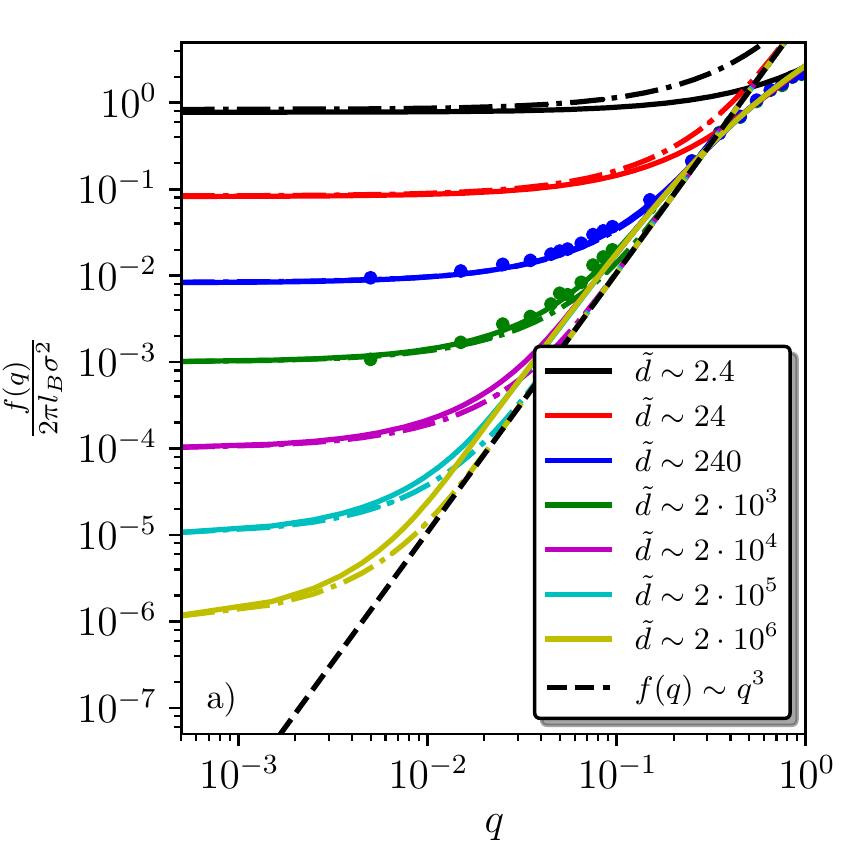}\includegraphics[width=0.5\textwidth,clip]{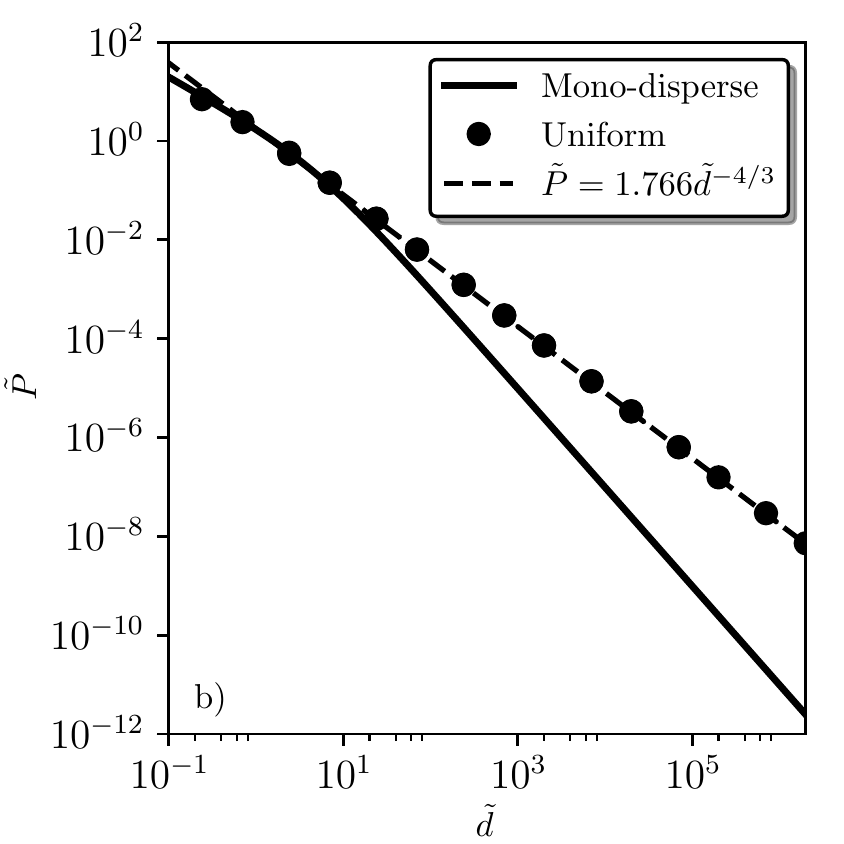}
\caption{Flat polydispersity situation ($\gamma=0$, $\alpha=3$), meaning that $n(q)$ is uniform in the interval $[0,1]$, for the two-plates situation.
a) Plot of the normalisation function $f(q)$, at different inter-plate distances. The circles are the estimates of $f(q)$ from the MC simulations. The solid lines show numerical
results from the poly-disperse PB treatment and finally, the dot-dashed lines are for the analytic results using Eq.~(\ref{fexp}). This is complemented by the dashed black line 
indicating the analytic asymptotic Eq.~(\ref{asympf}).
b) Equation of state. The black solid line is for the mono-disperse PB,  black circles show the numerical results for the poly-disperse PB and the black dashed line is for the 
analytic prediction Eq.~(\ref{analypPB}).}
\label{fig:fandPflat} 
\end{center}
\end{figure}

All results presented below are for the two-plates case. We start with 
the flat polydisperse distribution where $n(q)$ is a constant
(thus equal to $4\sigma$). Fig. \ref{fig:fandPflat} shows $f(q)$
and the pressure, predicted to behave at large $d$ as $d^{-4/3}$.
It can be seen that both PB resolutions and MC methods coincide,
and corroborate the analytical predictions. At any finite $d$,
the small $q$ limit of $f$ is finite, as given by Eq. (\ref{fexp}).

It can be seen that upon increasing $d$, $f$ evolves towards the
single plate behaviour $f \propto q^3$. It is also interesting
to note that whereas the mono- and polydisperse systems exhibit 
distinct pressure regimes at large $d$, they share very 
close pressure at smaller distances. This ``coincidence'' 
is made possible by the relevant choice of measuring distances
in unit of the Gouy length $(2\pi\ell_{\rm B}\sigma\langle q\rangle)^{-1}$
in both cases, but is otherwise all the less trivial as it also
holds beyond mean-field, at arbitrary Coulombic couplings \cite{TrST16}.
At large distances, the asymptotic prediction in $d^{-4/3}$ for the pressure is
well obeyed.

\section{Discussion}
\label{sec:discussion}

In the present section, we first summarise our main findings and present a more heuristic
derivation. This allows to generalise some of the two-plates results, that will then be
tested against both Poisson-Boltzmann numerical {solutions}, and Monte Carlo simulations.
We will also extend the analysis to a broader class of polydisperse distributions.
Finally, we address a central question, establishing the connection 
between our continuous mixture results, and the properties that characterise
discrete mixtures. Indeed, in any physically relevant system, $n(q)$ is discrete,
with the result that the minimum charge cannot vanish. Yet, 
physics is governed, at large distances,
by the small-$q$ features of $n(q)$, and more precisely, the new power-law regimes
reported in previous sections are ruled by the vicinity of $q=0$. This raises a legitimate concern,
and we explain in which sense the continuous limit is relevant to the discrete case.

\subsection{One-plate : summary of continuous distribution phenomenology}
Our treatment elaborates on the one-plate situation, screened by counterions only.
Some emphasis was put on the long-range behaviour, that is governed, expectedly,
by the population of counterions having the smallest valence ($q_{\text min}$).
When $q_{\text min}>0$, the system ultimately behaves like a monodisperse
one, having counterions of valence $q_{\text min}$. The one-plate density
thus behaves at large distances $x$ like $x^{-2}$ and likewise, the two-plate
pressure scales with distance $d$ like $d^{-2}$. Both function are furthermore independent of
the plate's bare charge $\sigma e$. 

The situation changes when polydispersity is considered. We have introduced an important
characteristics of polydispersity, through the exponent $\gamma$ specifying the low-$q$ 
behaviour of the valence distribution $n(q)$: $n(q) \propto \sigma q^\gamma$
for small $q$,
where the surface charge density $\sigma$ is kept for dimensional reasons.
We have $\gamma>-1$ to ensure normalisability. 
Decreasing $\gamma$ leads to an increase in the population of small $q$ counterions.
These are less sensitive to the electric field of the plate, that they consequently screen less.
Thus, the resulting one-plate electrostatic potential $\phi$ becomes longer range than in the monodisperse case,
and behaves (in absolute value) like $x^{1/(\gamma+3)}$. Formally, the monodisperse
case is recovered for $\gamma \to \infty$ (where the small $q$ regime is completely depleted),
for which our formula yields $\phi \propto x^0$, hinting at a logarithmic dependence.
For a given choice of index $\gamma$, we have shown that the counterionic number density $n(x)$
behaves (again at large $x$) like $x^{-2(\gamma+2)/(\gamma+3)}$, while the charge
density displays a different scaling: $\rho(x) \propto x^{-(2\gamma+5)/(\gamma+3)}$.
Again, when $\gamma \to \infty$, monodisperse phenomenology is recovered, with common asymptotic
dependences for $n$ and $\rho$ in $x^{-2}$. The fact that the power-law exponent is 
$\gamma$ dependent immediately implies that the saturation feature discussed in section
\ref{sec:onewall} is lost: when increasing $\sigma$, both $n(x)$ and $|\rho(x)|$ increase
without bound: $n(x) \propto \sigma^{2/(\gamma+3)}$ and $|\rho(x)| \propto \sigma^{1/(\gamma+3)}$.

\subsection{Heuristic derivation of two plates scaling laws, and comparison to numerical results}
The above one-plate considerations allow to recover some of our two-plates results,
and to generalise them beyond the case $\gamma=0$ that was worked out in detail
in section \ref{sec:twowalls}. We again focus on the large-distance asymptotic,
where in the vicinity of a given plate, the electrostatic potential is to a good approximation
provided by its one-plate limit and thus behaves like $x^{1/(\gamma+3)}$. For finite $\gamma$, 
the key to the large-$d$ physics is that there is always a population of counterions
that is too weakly charged to ``feel'' the electric potential. They have valence
$q$ smaller than some $d$-dependent threshold $q^*$, that we can simply estimate by the
following argument: $q^* \Delta \phi =1$, where $\Delta \phi$ is the potential
difference between the plate-contact, and the mid-plate point. Thus, we get the crossover
valence $q^* \propto d^{-1/(\gamma+3)}$. A relevant quantity is the total density
of the corresponding essentially ``free'' counterions, $n_f$ given by 
$\int_0^{q^*} n(q) dq \propto (q^*)^{1+\gamma} \propto d^{-(\gamma+1)/(\gamma+3)}$.
These ions are the main contributors to the force/pressure between the two plates;
having a surface density $n_f$ and a flat ($x$-independent) profile, their volume density
is simply given by $n_f/d$, a quantity that gives the inter-plate pressure.
We get here 
\begin{equation}
P \,\propto \, \frac{n_f}{d} \, \propto \, d^{-2(\gamma+2)/(\gamma+3)}.
\label{eq:Pheuristic}
\end{equation}
In the flat polydisperse $\gamma=0$ case, we recover the prediction $P \propto d^{-4/3}$
derived in section \ref{sec:twowalls}, and confirmed by PB and MC simulations.
Interestingly, we also retrieve the same functional
dependence for the inter-plate pressure as the one-wall number density
[same exponent $2(\gamma+2)/(\gamma+3)$, see Eq. (\ref{eq:densplateasymptotics})].
As a consequence, we can, along the same lines as in the monodisperse case,
define a non-linear dimensionless ratio $\cal R_\gamma$, by comparing the
true PB pressure at large $d$ to the superposition of the two one-plate densities
at $d/2$\footnote{
In the present symmetric two-wall setup, the PB pressure is simply given
by the mid-distance counterion density (up to a factor $kT$).}. We have shown above that ${\cal R}_\infty= \pi^2/8$
(monodisperse situation).
Computing $\cal R_\gamma$ requires the knowledge of all prefactors, which the 
present scaling analysis does not provide. Yet, the explicit results
of section \ref{sec:twowalls} for $\gamma=0$ yield 
${\cal R}_0 \simeq (3/2)^{2/3}\,  1.766 /2 \simeq 1.157$, slightly smaller than
${\cal R}_\infty$, but again larger than unity. Assuming that ${\cal R}_\gamma $ 
remains close to 1 for other values, this would mean that the error incurred by
computing the two-plate pressure at large $d$ from the superposition of
the one-plate densities, results in an underestimation, but not larger than 25\%.

\begin{figure}
\begin{center}
\includegraphics[width=0.5\textwidth,clip]{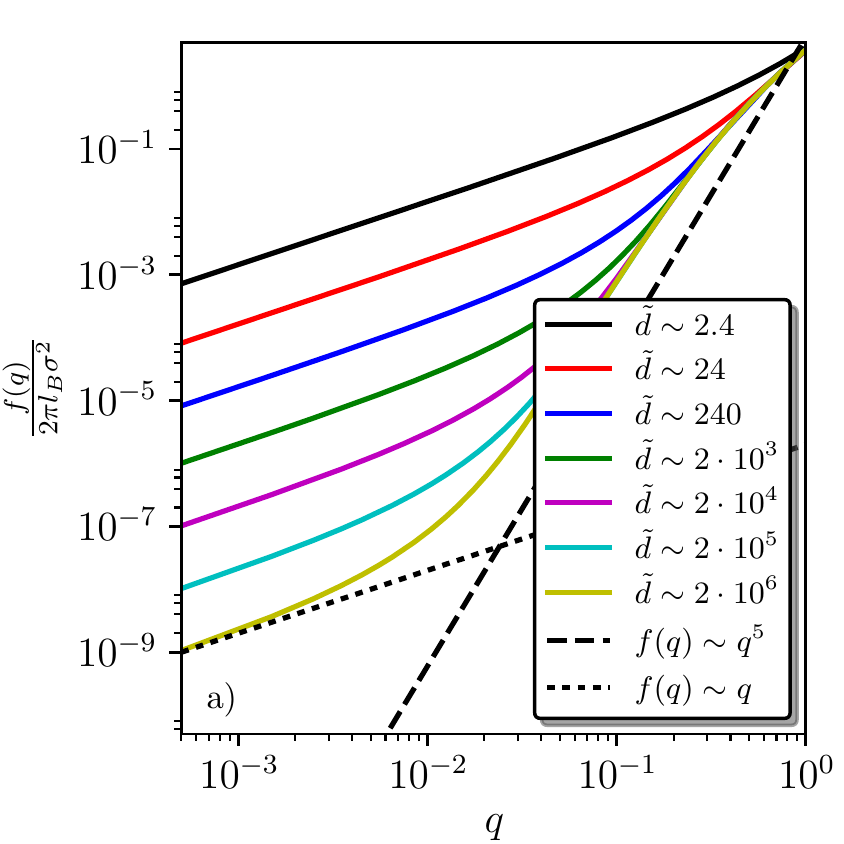}\includegraphics[width=0.5\textwidth,clip]{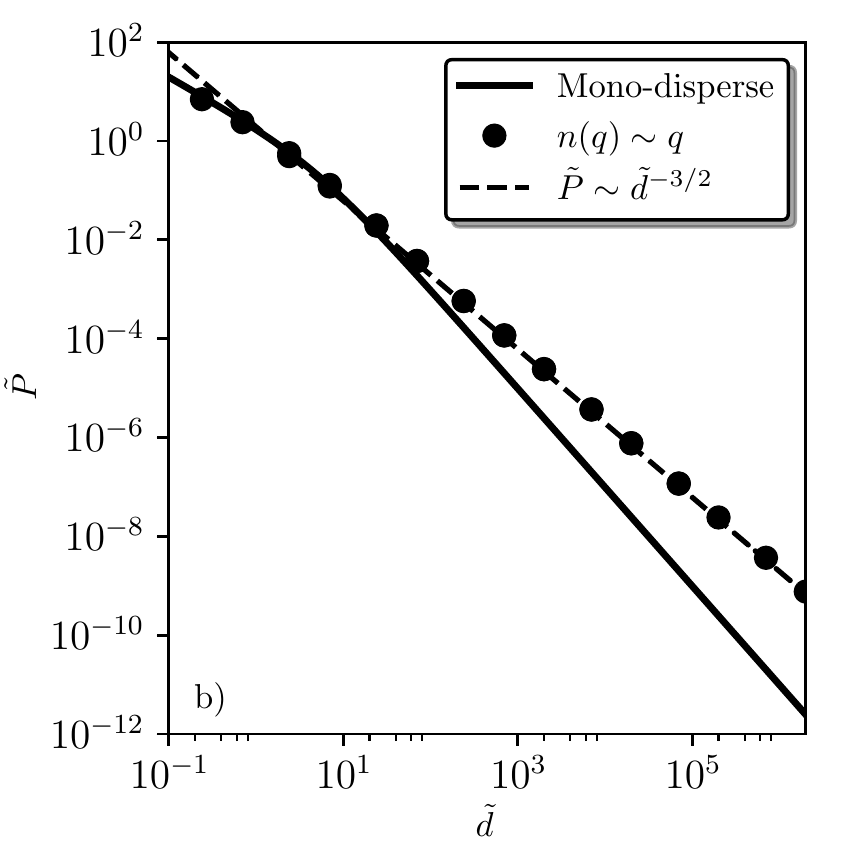}
\caption{{Skew distribution of counterions with $\gamma=1$,
meaning that small $q$ ions are less numerous than ions with a larger $q$
(small$-q$ depleted distribution.)}
a) Plot of the normalisation function $f(q)$, at different distances.
It can be seen that increasing $d$, $f$ adopts its one-plate shape 
in $q^{2\gamma+3}=q^5$ (see black dashed line), except below the threshold $q^*$ where
it shows the same behaviour as the parent $n(q)$ (here linear in $q$) (see black dotted line). (Colored lines) as Fig.~\ref{fig:fandPflat}.
b) Equation of state, which clearly shows a long-range $\tilde{d}^{-3/2}$-dependence,
as predicted by Eq. (\ref{eq:Pheuristic}).
The symbols and curves have the same meaning as in Fig. \ref{fig:fandPflat}.}
\label{fig:gamma1} 
\end{center}
\end{figure}

\begin{figure}
\begin{center}
\includegraphics[width=0.5\textwidth,clip]{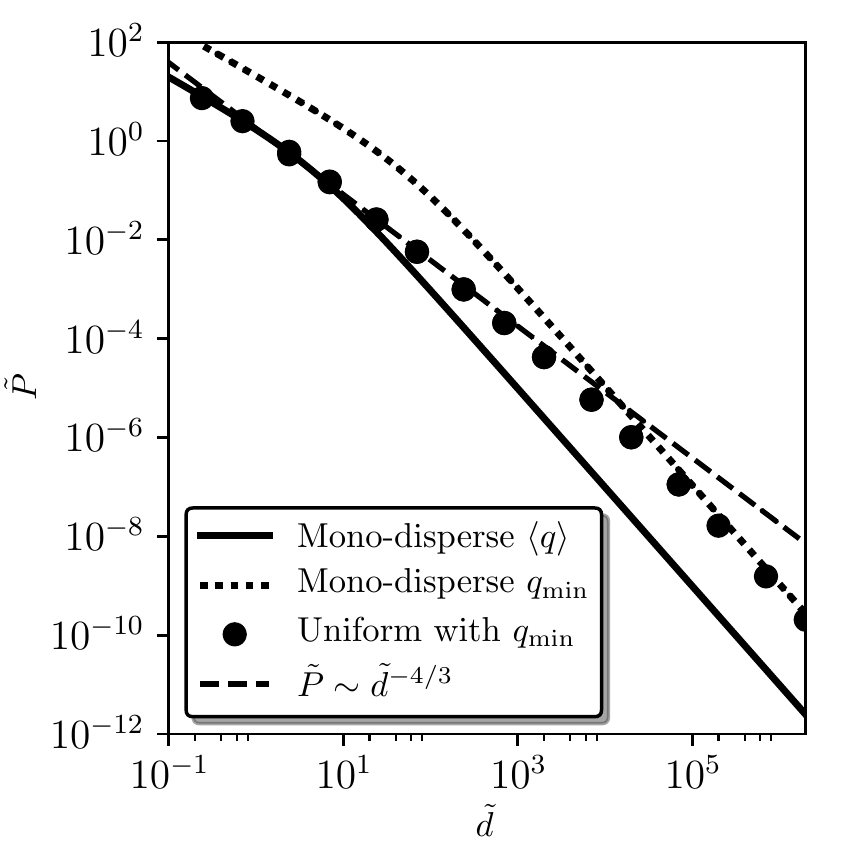}
\caption{Poisson-Boltzmann equation of state for a uniform $n(q)$ with $q$ in the range $[0.05,1]$.
{Black circles show the numerical results for the poly-disperse PB.} 
The two continuous curves show the PB monodisperse predictions, with two
distinct Gouy lengths: (black solid line) for $\mu^{-1}=2\pi\ell_{\rm B} \sigma \langle q\rangle$,
which is relevant at small distances,
and (black dotted line) for {$\mu^{-1}=2\pi\ell_{\rm B} \sigma q_{\text min}$}. These
two curves have an asymptotic $1/d^2$ decay.
An intermediate asymptotic with exponent $4/3$ sets in (dashed line).}
\label{fig:crossover} 
\end{center}
\end{figure}

\begin{figure}
\begin{center}
\includegraphics[width=0.5\textwidth,clip]{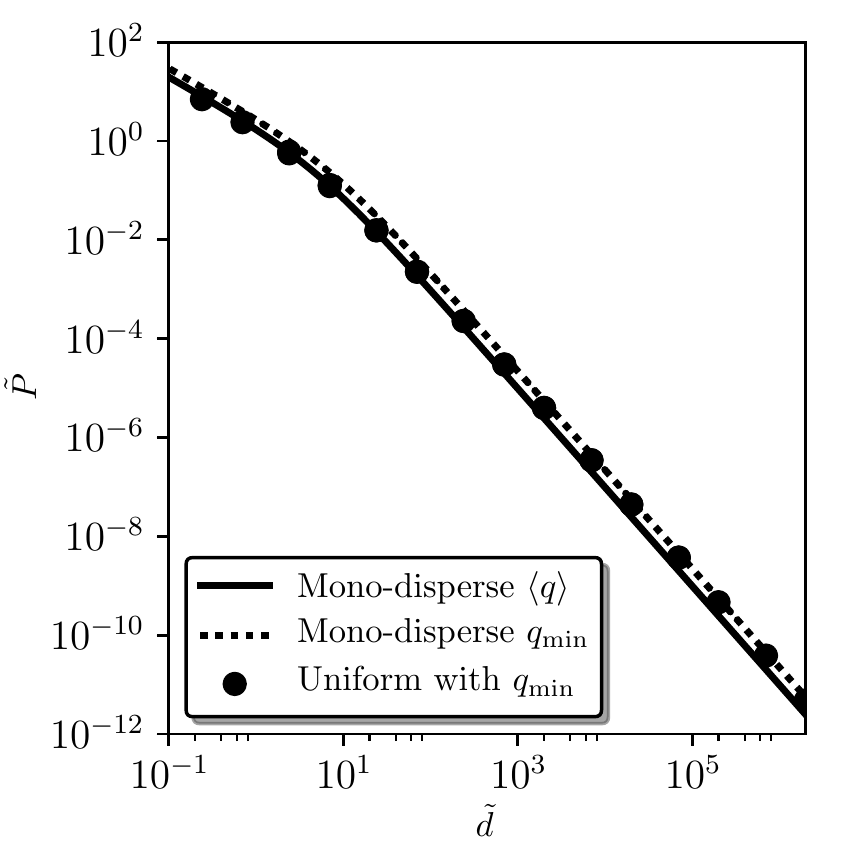}
\caption{Same as Fig. \ref{fig:crossover}, for a uniform $n(q)$ with $q$ in the range $[0.5,1]$
}
\label{fig:almostmono} 
\end{center}
\end{figure}

\begin{figure}
\begin{center}
\includegraphics[width=0.5\textwidth,clip]{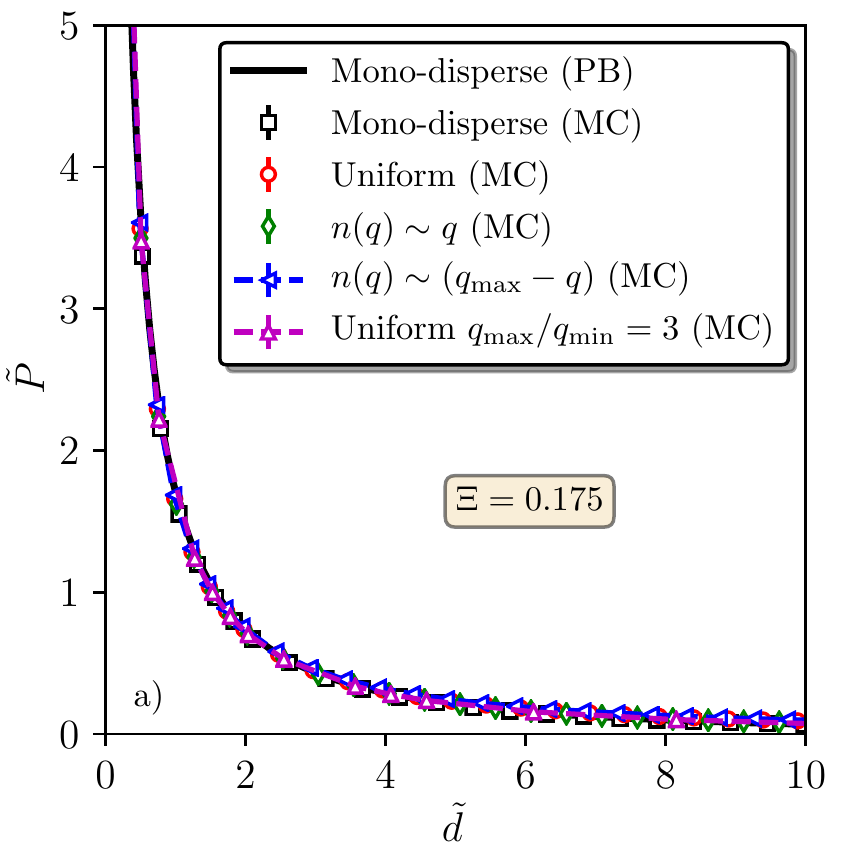}\includegraphics[width=0.5\textwidth,clip]{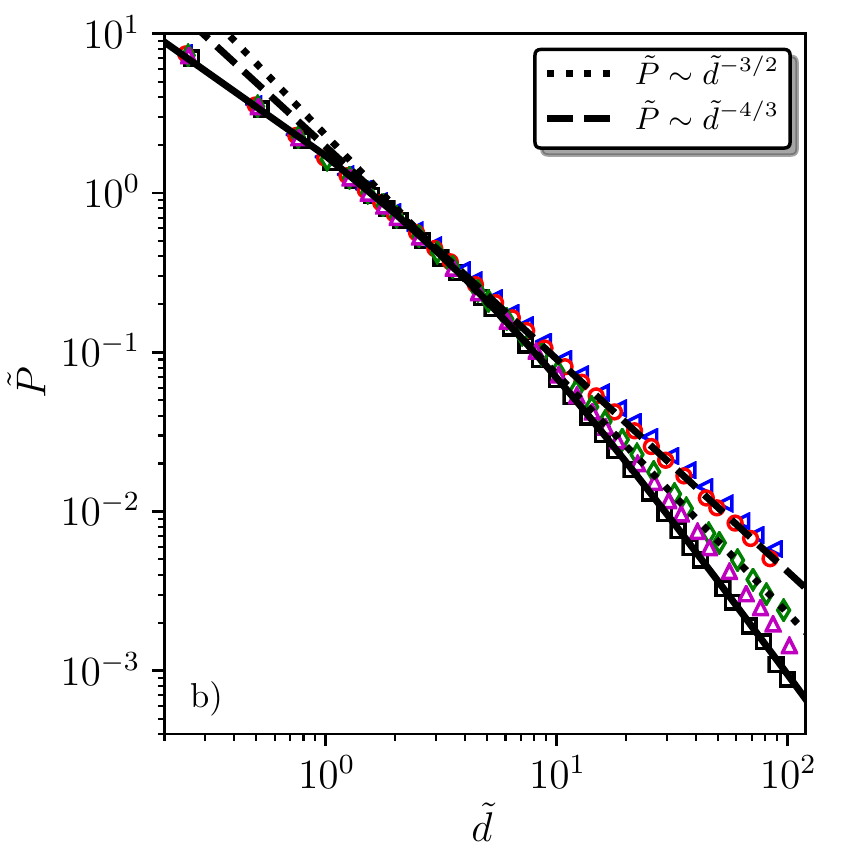}
\caption{Equation of state versus distance (where $\widetilde d$ is defined as
$2\pi\ell_{\rm B} \sigma \langle q\rangle \, d$)), for various valence
distributions: mono-disperse, uniform, skewed $n(q)\sim q$ (up to an upper cutoff), skewed $n(q)\sim (q_{\rm max}-q)$ (up to an upper cutoff), 
and uniform in $[q_{\rm min},q_{\rm max}]$ with $q_{\rm min}\neq 0$. The MC simulations are run at a coupling constant 
$\Xi=0.175$.
}
\label{fig:universality} 
\end{center}
\end{figure}

\begin{figure}[t!]
$$\includegraphics[width=0.5\textwidth,clip]{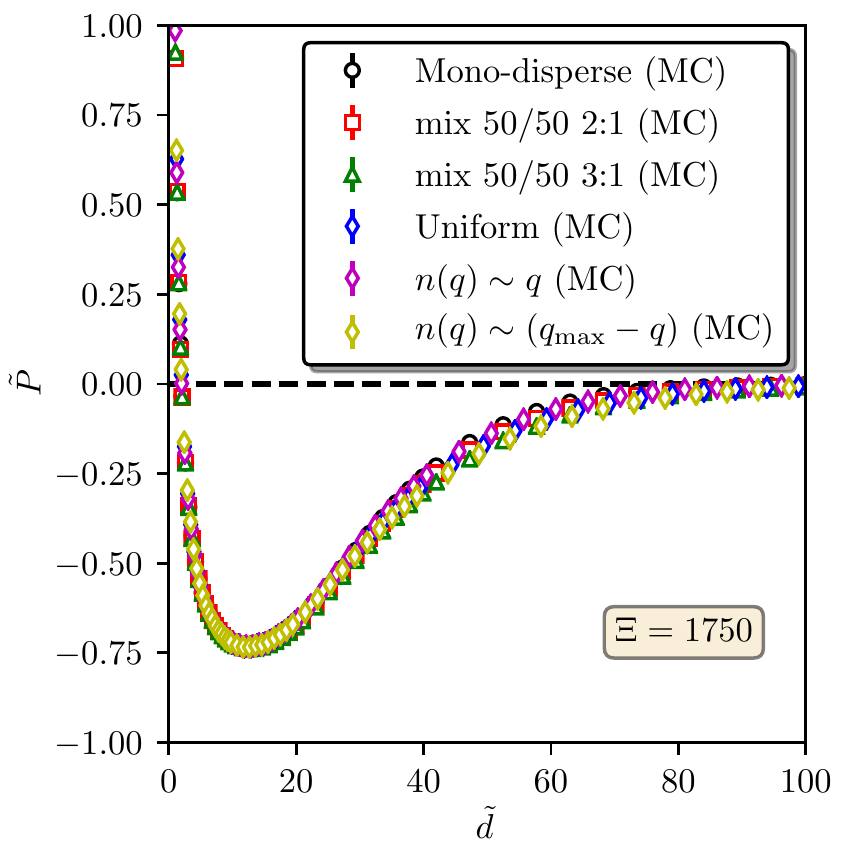}
\includegraphics[width=0.5\textwidth,clip]{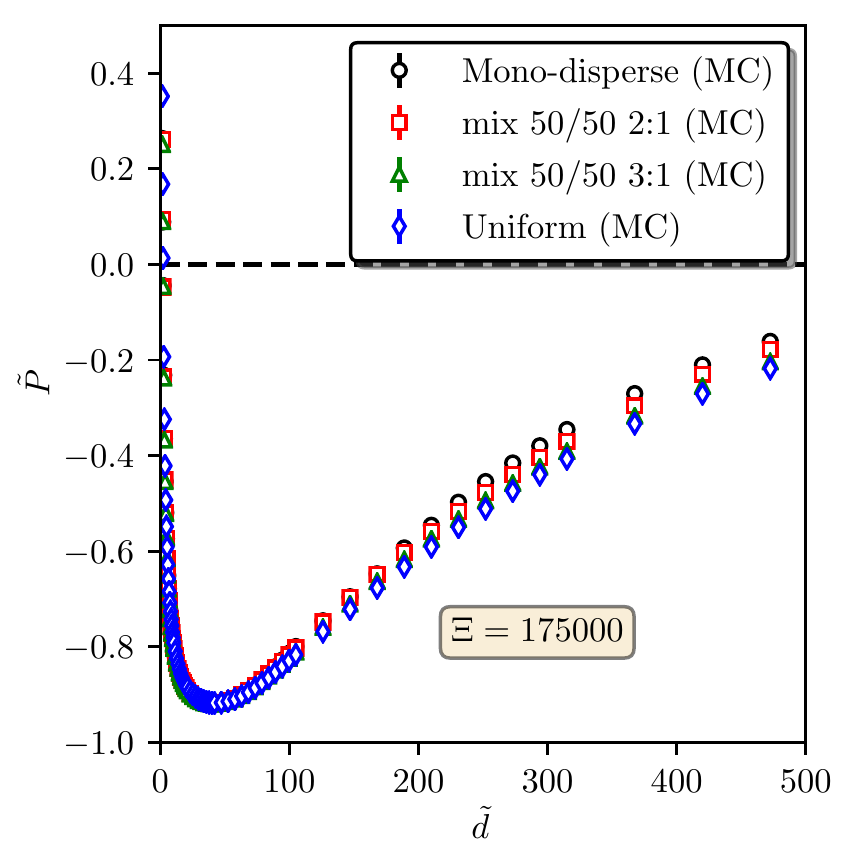}$$
\vspace{- 5 mm}
\caption{Monte Carlo reduced pressure versus normalized separation,
for strongly coupled systems ($\Xi=1750$ and $175000$) for which the Poisson-Boltzmann theory analysed in this paper
would completely fail. Significantly different distributions $n(q)$ are considered : monodisperse, bidisperse 
with $q_{\text min}/q_{\text max}=1/2$, $q_{\text min}/q_{\text max}=1/3$,
continuous ($q_{\text min}$=0, $\gamma=0$), and skewed (as in Fig. \ref{fig:universality}). }
\label{fig:sc}
\end{figure}

In Fig. \ref{fig:gamma1}, we show numerical PB results for $\gamma=1$, with thus
less small $q$ counterions than the $\gamma=0$ distribution discussed earlier. As a consequence,
the pressure exhibits a faster decay with $d$, predicted to be $d^{-3/2}$, see Eq. (\ref{eq:Pheuristic}). This is 
fully confirmed in Fig. \ref{fig:gamma1}. In addition, we have tested a number
of expectations, shaped on our previous analysis. First of all,
all distribution $n(q)$ having non-vanishing $n(0)$ should display the same
large-distance pressure, that of the $\gamma=0$ class. This was checked for the choice
$n(q)\propto (q_{\rm max}-q)$ (results not shown). Second, all distributions 
depleted near the origin ($n(q)=0$ for some range $q<q_{\text min}$)
should asymptotically behave like a monodisperse system, with counterion
valence $q_{\text min}$. Yet, if $q_{\text min}$ is not too large,
the system should require large distances $d$ before ``realising''
that $q_{\text min}$ is actually non vanishing. We should thus
expect a cross-over between the finite $\gamma$ behaviour in some
intermediate $d$-range, and the $\gamma=\infty$ ultimate decay.
This is what Fig. \ref{fig:crossover} clearly illustrates.
On the other hand, if $q_{\text min}$ and the maximum valence $q_{\text max}$ are not 
separated enough, the behaviour is of course close to its monodisperse
counterpart. Fig. \ref{fig:almostmono} shows that it is already
the case when $q_{\text max}/q_{\text min}=2$.
Finally, we show in Fig. \ref{fig:universality} that for quite a
large class of polydispersities, although the large-$d$ asymptotic may
be $n(q)$-dependent, the behaviour of pressure at smaller distances
is made rather universal, using properly scaled quantities \cite{TrST16}.
The data collapse reported is quite striking at short distances. 
For large $d$, the collapse is necessarily broken, since 
the different distributions studied correspond to distinct types, with various
$\gamma$ exponents. The corresponding decays range from $d^{-4/3}$ to $d^{-2}$,
including $d^{-3/2}$.
It is at this point relevant to stress that this collapse
holds beyond the mean-field regime which has been under scrutiny here,
as revealed in Fig.~\ref{fig:sc} {by Monte Carlo simulations for two strong coupling cases at various counterion mixtures.
The negative pressures seen at these high coupling parameters are a consequence of the now well-known ion-ion correlations, which is omitted in our mean-field
treatment. Surprisingly, these correlations do not break the collapse. 
These ion-ion correlations, that increase with the coupling parameter, can turn repulsive electrostatic 
interactions between two equally charged surfaces into
attractive ones.
Even though the chosen
coupling parameters are above realistic values for an aqueous electrolyte system 
(to be sure that the system is indeed dominated by ion-ion correlation effects), 
we emphasize that the same data-collapse holds for more experimentally relevant coupling parameters \cite{TrST16}. }
In addition, we address the possibility of a broader universality, including the tail
of the equation of state, in the following subsection.

\subsection{From continuous to discrete distribution of charges}
\label{ssec:transient}

So far, we only considered continuous distributions of charges $n(q)$, and we established the connection 
between the behaviour of $n(q)$ for $q \to 0$, and the long-range pressure or ionic profiles.
This may seem rather 
academic since any real physical system will exhibit some discreteness in $n(q)$. Our treatment thus raises
a two-pronged question: first, how ``close to continuous'' should a discrete $n(q)$ be to exhibit 
the predicted behaviour? Second, since the tail of the density profile (one-plate case),
or the long-distance equation of state (two plates) is necessarily ruled by the smallest charges 
in the system, how can the continuous power-laws derived for $q_{\text min}=0$ be observed 
in a discrete system having necessarily $q_{\text min}>0$? We note here that if a species
with a strictly vanishing charge is present in the mixture, it is simply discarded by the analytical
treatment worked out here.

\begin{figure}
\begin{center}
\includegraphics[width=0.5\textwidth,clip]{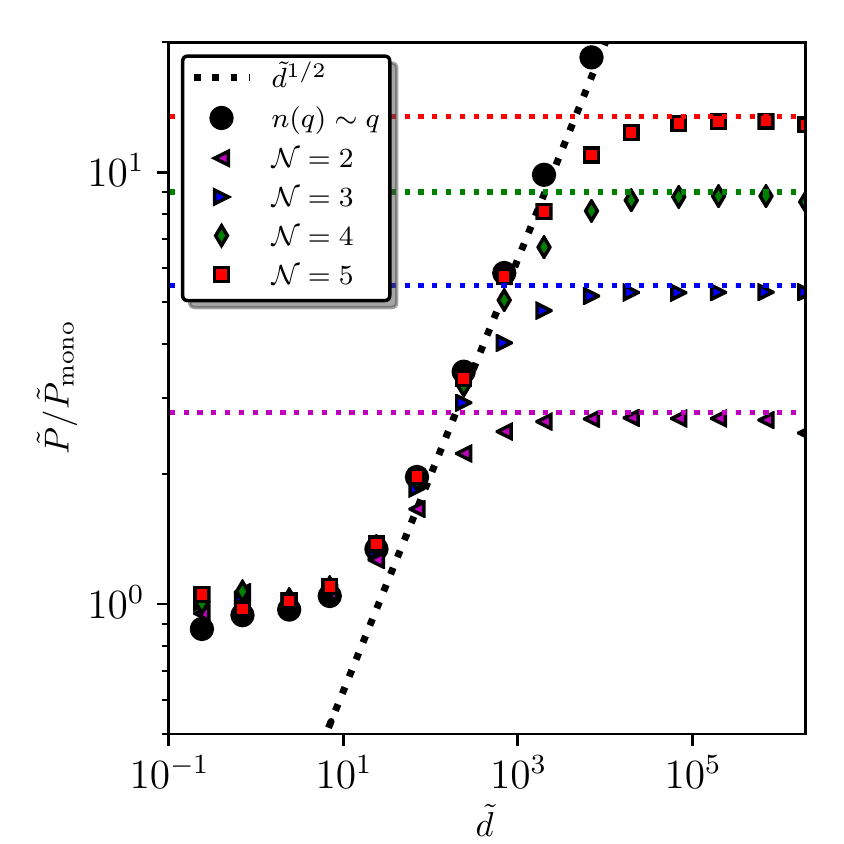}
\caption{Normalized pressure $\tilde{P}/\tilde{P}_{\text mono}$ {as obtained from the poly-disperse Poisson-Boltzmann}, where $\tilde{P}_{\text mono}$
is associated to a monodisperse system, with thus long-distance decay in $\tilde{d}^{-2}$. Five 
discrete $n(q)$ are considered with different number of species $\cal N$. 
The 1-2 system has charges 1/2 and 1 (${\cal N}=2 $, $q_{\text min}=1/2$),
the 1-3 system features charges 1/3, 2/3 and 1 (${\cal N}=3$, $q_{\text min}=1/3$) etc, and the 1-5 system
is with charges 1/5, 2/5,... 1. In all cases, the distribution is skewed, so that 
$n(q) \propto q$. As more and more species are added in the mixture, a transient asymptotic
sets {in} (dotted line), which exactly matches the continuous limiting distribution
(obtained for ${\cal N} \to \infty$, see the bullets). The value of the horizontal 
plateaus is given in the text. 
}
\label{fig:discreteskew} 
\end{center}
\end{figure}

Fig. \ref{fig:discreteskew} shows a rather striking result, establishing the
proximity between the discrete and continuous cases. The pressure arising in the Poisson-Boltzmann 
framework is computed for a number of mixtures, having $\cal N$ different species, with equispaced
charges {(like with a mixture of ions having integer charge values)}, and such that the number density of a constituent scales like $q$ itself (skewed distribution). 
This means that the index $\gamma$ introduced above is unity, and we expect a large distance
pressure in $d^{-3/2}$ for the continuous mixture with $q_{\text min}=0$ (see Eq. (\ref{eq:Pheuristic})). 
Since we show $P$ normalised by the monodisperse reference case, the continuous power-law for 
$P/P_{\text mono}$ should be in $d^{1/2}$ (dotted line), in good agreement with the 
disc symbols on the figure. On the other hand, all discrete mixtures should asymptotically 
behave, scaling-wise, like their monodisperse counterpart, at distances when only $q_{\text min}$ does remain
in the solution. This means that $P/P_{\text mono}$ in Fig. \ref{fig:discreteskew} is expected 
to flatten at large $d$, and converge towards a simple value: from our choice of units,
$\langle q \rangle^2/q_{\text min}^2$ (e.g., $9/4$ for ${\cal N}=2)$, in perfect agreement with the numerical data.
Yet, the most interesting feature is that for ${\cal N} =5$ already, the ``continuous'' power-law
is clearly visible, not asymptotically of course, but transiently (and for about a decade
in distance). Fig. \ref{fig:discreteskew} thereby shows how the continuous limit results
are recovered upon increasing ${\cal N}$, and that an arguably small value of ${\cal N}$ is
sufficient to exhibit some of the hallmarks of continuous systems.

\begin{figure}
\begin{center}
\includegraphics[width=0.5\textwidth,clip]{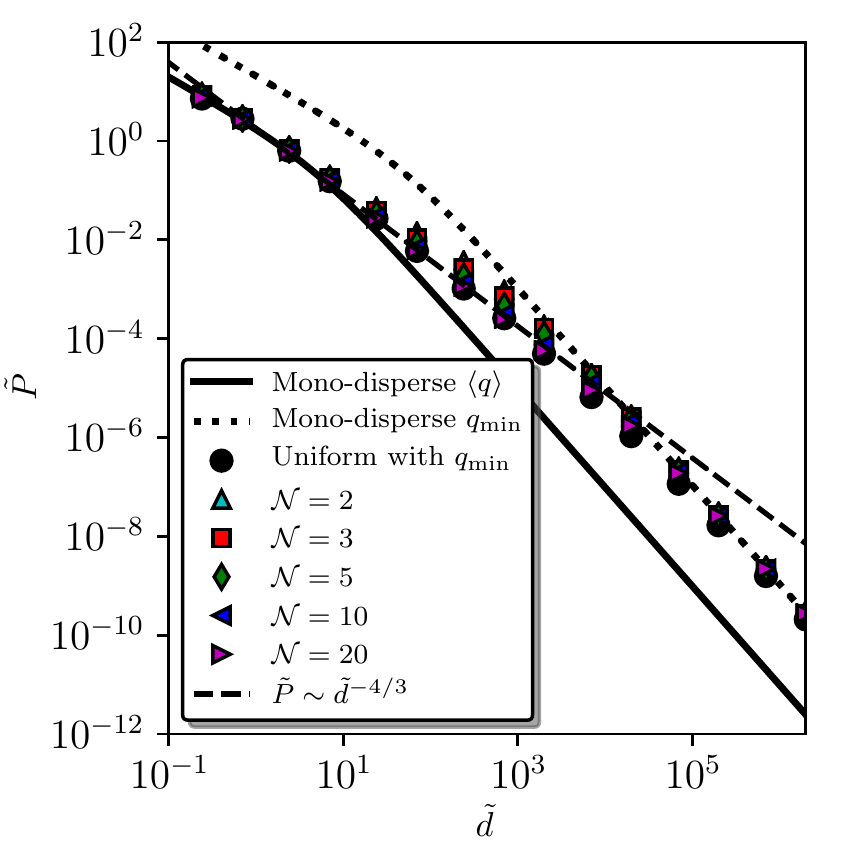}
\caption{Illustration of quasi-universality for different distributions
exhibiting the same value of $q_{min}$. Five discrete distributions are considered,
with equidistributed charges starting at $q_{\text min}=0.05$ and such that $\langle q \rangle =1/2$,
and ${\cal N}=2$, 3, 5, 10 and 20 charges.
The pressure is compared to those of the corresponding continuous model ${\cal N} \to \infty$,
and of the two limiting monodisperse regimes: one with $n(q) = \delta(q-1/2)$ (lower bound shown by the
continuous line),
and the other for $n(q) = \delta(q-0.05)$ (upper bound, dotted line).
}
\label{fig:discreteuniv} 
\end{center}
\end{figure}

Fig. \ref{fig:discreteuniv} illustrates a similar effect, with the distinction that 
all distributions shown share the same value of $q_{\text min}$, even the continuous 
case. Here, we chose a ``flat'' situation where $n(q)$ is the same for all $q$ values.
The first message conveyed is that all curves are reasonably close (and all the closer as we are displaying data
on a log scale), so that discreteness effects are not paramount. While the ${\cal N}=20$ charges case 
is arguably close to the continuous limit, considering   ${\cal N}=2-3$ peaks is already sufficient
to observe the main trend. The second message pertains to the transient asymptotic.
In Fig. \ref{fig:discreteuniv}, the dashed line with slope $-4/3$ is the prediction derived in this 
work (corresponding to $\gamma=0$ and $q_{\text min}=0$). 
While all distributions yield a large $d$ tail in $d^{-2}$ since $q_{\text min} \neq 0$,
the ``continuous/$q_{\text min}=0$'' power-law in $d^{-4/3}$ does hold 
approximately in a finite distance range, over 4 decades.
The salient features of Fig. \ref{fig:discreteskew} and \ref{fig:discreteuniv} explain 
why the analytical derivations proposed here have relevance
for discrete systems as well.

\acknowledgments
The support received from the Grant VEGA No. 2/0015/15 is acknowledged.



\appendix
\section{One-plate geometry : long-range features}
\label{app:asymptotic}

In this appendix, we establish the connection between the small $q$-behaviour 
of function $f(q)$ as encoded in Eq. (\ref{fqasymp}), with the long-distance 
regime of densities (charge, and number densities, that do differ in general).
Injecting (\ref{fqasymp}) into (\ref{rewrite}), we get
\begin{equation} \label{asympt}
\left[ \phi'(\widetilde{x}) \right]^2 = \frac{4}{\langle q\rangle^2} 
a\Gamma(\alpha+1) \frac{1}{[-\phi(\widetilde{x})]^{\alpha+1}} , 
\qquad \widetilde{x}\to\infty ,
\end{equation}
where $\Gamma$ denotes the Gamma function. 
The solution of this asymptotic equation is searched in the form
\begin{equation} \label{asymppot}
\phi(\widetilde{x}) \mathop{\sim}_{\widetilde{x}\to\infty} 
- b \left( \frac{\widetilde{x}}{\langle q\rangle} \right)^{\beta} .
\end{equation}
Inserting this ansatz into Eq. (\ref{asympt}), the exponent $\beta$ and
the prefactor $b$ are determined self-consistently as
\begin{equation} \label{betab}
\beta = \frac{2}{\alpha + 3} , \qquad 
b^{\alpha+3} = a \Gamma(\alpha+1) (\alpha+3)^2 .
\end{equation}
The large-distance behaviour of the electric potential reads
\begin{equation} \label{phiasym}
\phi(\widetilde{x}) \mathop{\sim}_{\tilde{x}\to\infty} - 
\left[ a \Gamma(\alpha+1) (\alpha+3)^2 \right]^{\frac{1}{\alpha+3}} 
(\widetilde{x}/\langle q\rangle)^{\frac{2}{\alpha+3}} .
\end{equation}
The logarithmic dependence found in the monodisperse case for $\phi$ changes to
an asymptotic power-law behaviour with non-universal index and prefactor,
depending on the model's parameters $a$ and $\alpha$. This is 
the consequence of a less efficient screening with counterions having
a small $q$. As we shall see below, large-$\beta$ values correspond to 
systems with enhanced population with $q$ near 0, with resulting impeded
screening. 
The asymptotic number density profile of particles reads as
\begin{eqnarray} 
n(\widetilde{x}) & = & \int_0^1 {\rm d}q\, f(q) {\rm e}^{q\phi(\tilde{x})} 
\mathop{\sim}_{\tilde{x}\to\infty} 2\pi \ell_{\rm B} \sigma^2 a 
\frac{\Gamma(\alpha+1)}{[-\phi(\widetilde{x})]^{\alpha+1}} \nonumber \\ 
& = & 2\pi \ell_{\rm B} \sigma^2 \frac{[a\Gamma(\alpha+1)]^{\frac{2}{\alpha+3}}}{
(\alpha+3)^{2\left(\frac{\alpha+1}{\alpha+3}\right)}}
\frac{1}{(\widetilde{x}/\langle q\rangle)^{2\left(\frac{\alpha+1}{\alpha+3}\right)}} . 
\label{nasymp}
\end{eqnarray}    
Similarly, the asymptotic charge density profile reads as
\begin{eqnarray}
\frac{\rho(\widetilde{x})}{(-e)} 
& = & \int_0^1 {\rm d}q\, q f(q) {\rm e}^{q\phi(\tilde{x})} 
\mathop{\sim}_{\widetilde{x}\to\infty} 2\pi \ell_{\rm B} \sigma^2 a 
\frac{\Gamma(\alpha+2)}{[-\phi(\widetilde{x})]^{\alpha+2}} \nonumber \\
& = & 2\pi \ell_{\rm B} \sigma^2
\frac{[a\Gamma(\alpha+1)]^{\frac{1}{\alpha+3}}(\alpha+1)}{
(\alpha+3)^{2\left(\frac{\alpha+2}{\alpha+3}\right)}}
\frac{1}{(\widetilde{x}/\langle q\rangle)^{2\left(\frac{\alpha+2}{\alpha+3}\right)}} . 
\label{rhoasymp}
\end{eqnarray}    
It is easy to check that these asymptotic behaviours fulfil the exact 
relation $\rho(\widetilde{x})/(-e) = n'(\widetilde{x})/\phi'(\widetilde{x})$,
see Eq. (\ref{charge}).
We conclude that the non-universal large-$x$ behaviour of 
the reduced potential, the number and charge density profiles are determined 
by the small-$q$ behaviour of the normalisation function $f(q)$. This was
expected, since those counterions with the smallest $q$ are the least sensitive to 
the created electric field, and thus the most delocalised. 

Let us rewrite the $n-f$ relation (\ref{nfself}) in terms of the dimensionless
$\widetilde{x}$, 
\begin{equation} \label{nq}
\frac{n(q)}{\sigma} = \frac{f(q)}{2\pi\ell_{\rm B}\sigma^2} 
\int_0^{\infty} \frac{{\rm d}\widetilde{x}}{\langle q\rangle}\, 
{\rm e}^{q\phi(\tilde{x})} .
\end{equation}
In the limit $q\to 0$, we can use the small-$q$ asymptotic (\ref{fqasymp})
in Eq. (\ref{nq}) to write down
\begin{equation} \label{smallq}
\frac{n(q)}{\sigma} \mathop{\sim}_{q\to 0} a q^{\alpha}
\int_0^{\infty} \frac{{\rm d}\widetilde{x}}{\langle q\rangle}\, 
{\rm e}^{q\phi(\tilde{x})} .
\end{equation}
The integral on the rhs of this equation diverges as $q\to 0$
due to the integration of unity over an infinite support.
We do not know the functional form of the reduced potential $\phi$ 
at small $\widetilde{x}$, but we do know its asymptotic form (\ref{phiasym}) 
at large $\widetilde{x}$.
Since the integral diverges, any integration on a finite interval
does not affect the leading divergent term.
Based on this fact we make an assumption which will be later verified 
numerically on a specific model: to study the small-$q$ divergence of 
the integral in (\ref{smallq}) it is sufficient to insert there 
the asymptotic large-$\widetilde{x}$ formula for the potential (\ref{phiasym}).
If this assumption is correct, we obtain
\begin{equation}
\int_0^{\infty} \frac{{\rm d}\widetilde{x}}{\langle q\rangle}\,  
{\rm e}^{q\phi(\tilde{x})}
\mathop{\sim}_{q\to 0} \frac{1}{2} \frac{1}{\sqrt{a\Gamma(\alpha+1)}}
\Gamma\left( \frac{\alpha+3}{2} \right) q^{-\frac{\alpha+3}{2}} .
\end{equation}
Consequently,
\begin{equation} \label{finalrel}
\frac{n(q)}{\sigma} \mathop{\sim}_{q\to 0} \frac{1}{2} 
\sqrt{\frac{a}{\Gamma(\alpha+1)}}
\Gamma\left( \frac{\alpha+3}{2} \right) q^{\frac{\alpha-3}{2}} .
\end{equation}
We see that in the $q\to 0$ limit, the density distribution goes to a nonzero 
constant when $\alpha=3$, it vanishes when $\alpha>3$ and diverges
for $\alpha<3$.
Since the surface density of particles must be finite, the density distribution 
should be integrable for small $q$ and we have the restriction $\alpha>1$.

The crucial relation (\ref{finalrel}) relates the small-$q$ behaviour
of the density function of particles $n(q)$, which is given from the outset
in the direct formulation of the problem, to the small-$q$ behaviour of 
the normalisation function $f(q)$ (\ref{fqasymp}). 
It turns convenient to introduce a parameter $\gamma$ through 
\begin{equation}
\alpha = 2\gamma + 3 , \qquad a = c^2 \frac{2^{2\gamma+5}}{\sqrt{\pi}}
\frac{\Gamma\left( \gamma+\frac{5}{2} \right)}{(\gamma+2) \Gamma(\gamma+3)} .
\end{equation}
Indeed, $\gamma$ characterises the behaviour of $n(q)$ at small $q$, which is physically more relevant 
that the behaviour of $f(q)$, see the main text.

\section{Computational aspects}
\label{app:num}
\subsection{Poisson-Boltzmann resolution}
The polydisperse Poisson-Boltzmann equation, Eq.~(\ref{intPBpoly1}), was solved numerically through a real-valued variable-coefficient ordinary differential equation (ODE) 
solver with an initial guess of $f_g(q)$-distribution aimed to target a particular $n_t(q)$-distribution.
For each such $f_g(q)$ guess, a new corresponding $n_g(q)$-distribution is found through Eq.~(\ref{rel}). 
A new guess for the correct $f_t(q)$-distribution is then generated by a mixing of the new 
distribution $f_{g,{\rm new}}(q)$ with the old one, $f_{g,{\rm old}}(q)$. The new $f_{g,{\rm new}}(q)$-distribution is found
from a re-distribution of the old $f_{g,{\rm old}}(q)$ through
\begin{equation}
f_{g,{\rm new}}(q) = f_{g,{\rm old}}(q) \frac{n_t(q)}{n_g(q)}.
\end{equation}
Mixing of $f_{g,{\rm old}}(q)$ and $f_{g,{\rm new}}(q)$ is then done with a small fraction of the new guess compared to the old.
However, such a mixing of $f(q)$ runs into the risk of creating unrealistic negative values of pressures and imaginary electrostatic potentials 
(see \textit{e.g.} Eq.~(\ref{pressurepolyi})). 
To avoid such negative pressures a renormalisation of the total distribution $f(q)$ 
is performed using Eq.~(\ref{pressurepolyi}) such that the pressure at contact matches the pressure calculated from the mid-plane.
Such a scheme usually reaches a  convergence just after a few iterations.
Consistency was then checked by calculating the pressure through the two pressure routes, at contact and across the mid-plane according to
Eq.~(\ref{pressurepolyi}). 

Alternatively one can solve the second order ODE, instead of the redefined first order ODE, for the poly-disperse case according to Eq.~(\ref{PB1}), but at the expense of 
time to convergence. Both routes yield however the same results. 
A typical calculation was based on a discretisation of $q$, $n(q)$, and $f(q)$ into 1000 bins as well as discretisation of the $x$ axis (usually by some
fractions of $\ell_{\rm B}$). 
We verified that our solutions did not depend on these discretizations/binnings by increasing or decreasing the number of bins/steps.




\subsection{Monte Carlo simulations}
We have performed Monte-Carlo simulations in a quasi-2D geometry. Long-ranged electrostatic interactions are handled with
Ewald summation techniques corrected for quasi-2D-dimensionality by introducing a vacuum
slab in the $z$-direction perpendicular to the surfaces \cite{Berkowitz,Mazars}. We verified that our vacuum slab is sufficiently wide, so as not to influence the results. 
All simulations consisted of 512 point charges while the surfaces are modelled as structureless infinite plates with uniform surface
charge densities equal to $\sigma e$.
Simulations were performed both for discrete mixtures of charges as well
as for quasi-continuous\footnote{
Quasi in the sense that we have a finite number of ions. These continuous distributions
are generated by randomly assigning charges according to the desired $n(q)$-distribution.} distributions of charges, $q\in [q_{\rm min},q_{\rm max}]$. Standard displacement trials were performed with an acceptance ratio of around 30\%. 
Pressures were estimated using
the contact densities and the contact theorem as well as across the mid-plane, and were collected over $10^5$ Monte Carlo cycles. These two approaches yielded the same
pressures within statistical noise/errors. Estimates of $f(q)$ for each mixture was done by measuring the contact values at the wall
for each $q$-values (via a discretisation). 
To be able to compare with our Poisson-Boltzmann calculations, we have performed the simulations at sufficiently low coupling parameter {($\Xi=0.175$).
To show quasi-universality also beyond mean-field we have performed simulations at higher coupling parameters ($\Xi=1750$ and $\Xi=175000$).}



\end{document}